\documentclass[showpacs,preprintnumbers,twocolumn,amsmath,amssymb,superscriptaddress]{revtex4}
\usepackage{graphicx}
\usepackage{dcolumn}
\usepackage{bm}

\begin{document}

\title{Microscopic description of quadrupole collectivity in neutron-rich nuclei across the  N=126 
shell closure}

\author{R. Rodr\'{\i}guez-Guzm\'an}
\email{raynerrobertorodriguez@gmail.com}
\affiliation{Physics Department, Kuwait University, Kuwait 13060}

\author{L.M. Robledo}
\email{luis.robledo@uam.es}
\affiliation{Departamento  de F\'{\i}sica Te\'orica, M\'odulo 15,
Universidad Aut\'onoma de Madrid, 28049-Madrid, Spain}

\author{M. M. Sharma}
\email{madan.sharma@ku.edu.kw}
\affiliation{Physics Department, Kuwait University, Kuwait 13060}

\date{\today}

\begin{abstract}	
The quadrupole collectivity  in  Nd, Sm, Gd, Dy, Er, Yb, Hf and W 
nuclei with neutron numbers 122 $\le$ N $\le$ 156 is studied, both at 
the mean field level and beyond, using the Gogny energy density 
functional. Besides the robustness of the N=126 neutron shell closure, 
it is shown that the onset of static deformations in those isotopic 
chains with increasing neutron number leads to an enhanced stability 
and further extends the corresponding two-neutron driplines far beyond 
what could be expected from spherical calculations. Independence of the 
mean field predictions with respect to the particular version of the 
Gogny energy density functional employed is demonstrated  by comparing 
results based on the D1S and D1M parameter sets. Correlations beyond 
mean field  are taken into account in the framework of the  angular 
momentum projected generator coordinate method calculation. It is shown 
that N=126 remains a robust neutron magic number when dynamical effects 
are included. The analysis of the collective wave functions, average 
deformations and excitation energies indicate that, with increasing 
neutron number, the zero-point quantum corrections lead to dominant 
prolate configurations in the 0$_{1}^{+}$, 0$_{2}^{+}$, 2$_{1}^{+}$ and 
2$_{2}^{+}$ states of the studied nuclei. Moreover, those dynamical 
deformation effects provide an enhanced stability that  further 
supports  the mean field predictions, corroborating a shift of the 
r-process path to higher neutron numbers. Beyond mean field 
calculations provide a smaller shell gap at N=126 than the mean field 
one in good agreement with previous theoretical studies. However, the 
shell gap still remains strong enough in the two-neutron driplines.
\end{abstract}


\maketitle

%

\section{Introduction}

The evolution of the shell structure and the associated quadrupole 
collectivity with nucleon number is among the most prominent features 
in atomic nuclei. Its study has received renewed interest in recent 
years due to the progress in our understanding of neutron-rich nuclei 
far away from the valley of stability brought by the Radioactive Ion 
Beam (RIB) facilities set up all over the world. Shell effects in 
neutron-rich nuclei turn out to be  quite challenging and, at least in 
some cases, they cannot be interpreted using the experience accumulated 
for stable systems \cite{Sorlin-review}. A typical example is the 
weakening/erosion of the N=20 and N=28 magic numbers in light 
neutron-rich nuclei (see, for example, 
\cite{Michimasa,Thibault,Motobayashi,Detraz,Guillemaud-Mueller,Bastin,Takeuchi,Caurier,Utsuno,NPA-2002,PLB-Rayner-Egido-Robledo,RaynerN20-PRC2000,RaynerNe30,Peru-2028,Sharma32Mg,Terasaki,Stoisov,SM-28,RaynerN28,meanfieldN28-1,meanfieldN28-2,meanfieldN28-3,Lewi-N28,Sorlin-N28,Scheit-N28,Glas-N28} 
and references therein). 

In a mean field framework, the nucleus $^{32}$Mg is predicted to have a 
spherical ground state 
\cite{NPA-2002,PLB-Rayner-Egido-Robledo,RaynerN20-PRC2000,Sharma32Mg,Terasaki,Stoisov}. 
However, the experimental B(E2, 0$_{1}^{+}$ $\rightarrow$ 2$_{1}^{+}$) 
value \cite{Motobayashi}, the excitation energy of the 2$_{1}^{+}$ 
state \cite{Detraz,Guillemaud-Mueller} as well as the ratio 
$E_{4_{1}^{+}}/E_{2_{1}^{+}}$=2.6 are all consistent with those of a 
deformed ground state. Within the Shell Model, the increased quadrupole 
collectivity  in nuclei of this region has been explained by invoking 
neutron excitations across the N=20 shell gap \cite{Caurier,Utsuno}. On 
the other hand, it has been shown that  effects not explicitly taken 
into account at the mean field level like the restoration of 
broken symmetries and configuration mixing, can account for a deformed 
ground state in $^{32}$Mg and $^{30}$Ne  as well as the main physical 
trends in nuclei around the {\it{island of inversion}} 
\cite{NPA-2002,PLB-Rayner-Egido-Robledo,RaynerN20-PRC2000,RaynerNe30,Peru-2028}. 
This has been further corroborated by the recent study of the 
quadrupole collectivity in $^{28,30}$Ne and $^{34,36}$Mg 
\cite{Michimasa}. The situation is slightly different in the case of 
neutron-rich nuclei around N=28 where Shell Model \cite{SM-28},  mean 
field 
\cite{NPA-2002,RaynerN28,meanfieldN28-1,meanfieldN28-2,meanfieldN28-3} 
and beyond mean field \cite{NPA-2002,Peru-2028,RaynerN28} calculations 
predict deformed ground states, in agreement 
with experimental results 
\cite{Lewi-N28,Sorlin-N28,Scheit-N28,Glas-N28}, and indicating  that this 
neutron shell closure is more fragile than the N=20 one.

%
%
\begin{figure*}
\includegraphics[width=1.0\textwidth]{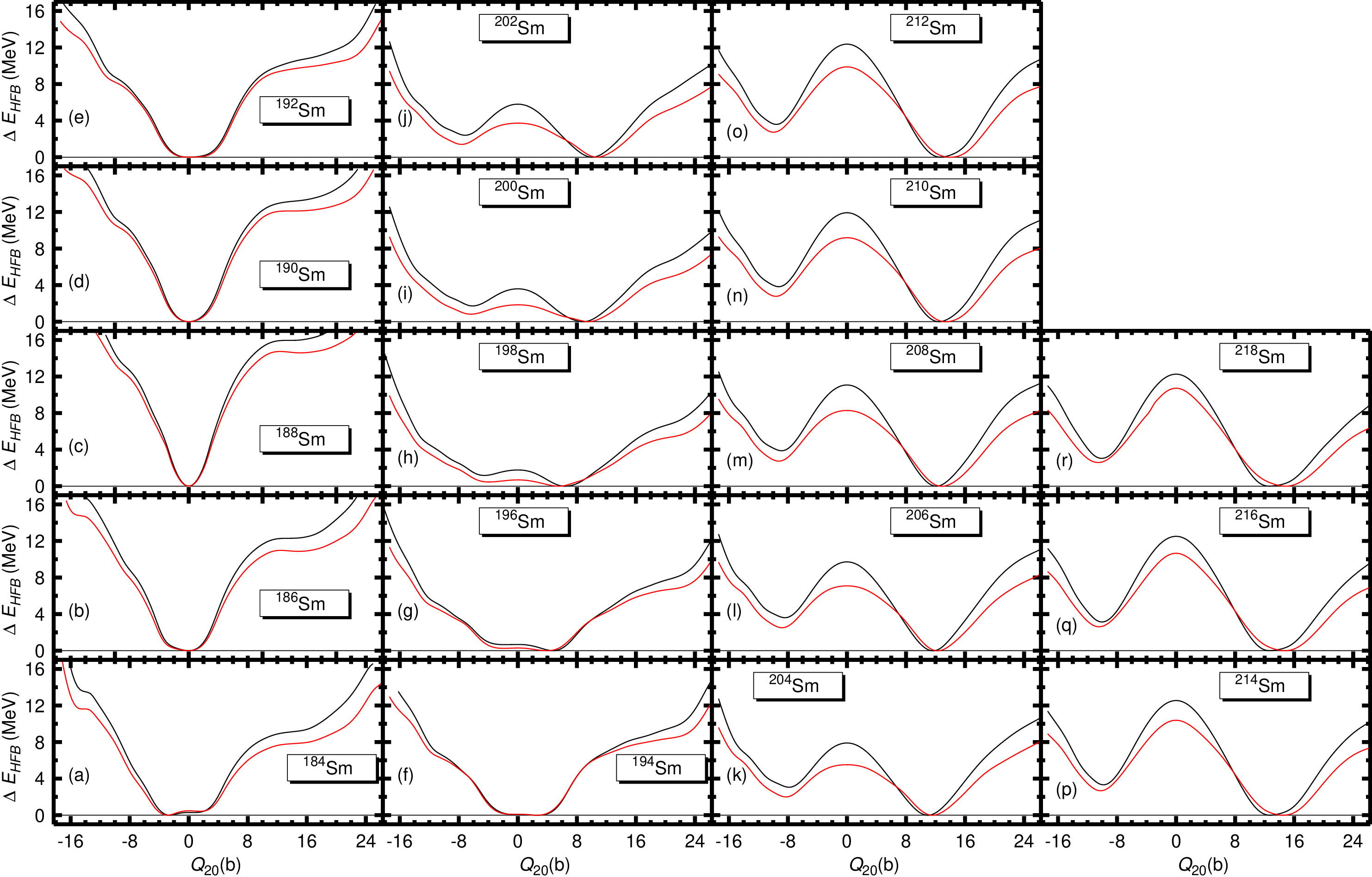}
\caption{ (Color online) In panels (a)-(r) the MFPECs obtained 
for the $^{184-218}$Sm isotopes
with the Gogny-D1M  (black) and Gogny-D1S (red) EDFs are depicted.  
Each curve is referred to its absolute minimum.  
Calculations have been carried out with M$_{z,Max}$=14.
For more details, see the main text.
}
\label{fig-PEC-Sm_D1M}  
\end{figure*}

The previous examples already illustrate the challenges encountered in 
the theoretical description of neutron-rich nuclei. They  suggest that 
caution must be taken when invoking the weakening/erosion of neutron 
magic numbers and  reveal that  plain mean field approximations 
\cite{rs}, while valuable as a starting point, may not be sufficient 
since the quadrupole properties in the considered nuclei may actually 
be determined  by the subtle interplay between quantum corrections 
stemming from the restoration of the broken symmetries (mainly the 
rotational one) and fluctuations in the collective degrees of freedom. 
Similar conclusions can be extracted from other beyond mean field 
calculations (see, for example, 
\cite{PERU-MARTINI,Vetrenar-BYMF-1,Vetrenar-BYMF-2} and references 
therein).  On the neutron deficient side, symmetry-projected 
configuration mixing results indicate that Z=82 remains, on the 
average, as a conserved magic proton number for Pb isotopes and support 
the experimental evidences for rotational bands built on coexisting 
low-lying  prolate and oblate 0$^{+}$ states 
\cite{Rayner-Pb,Duguet,Bender-Pb-1}.

Shell effects play a crucial role in understanding the nucleosynthesis 
of nuclei heavier than Fe via the r-process 
\cite{r-process-4,r-process-ref1,r-process-ref2,r-process-ref3}. 
Though different r-process scenarios and uncertainties are still 
under debate (see, for example, \cite{r-Wanajo}), it is commonly accepted that 
the r-process passes through the neutron numbers N=50, 82 and 126 and 
that the synthesis of nuclei around them is revealed in the peaks of 
the r-process abundances around the mass numbers A $\approx$ 80, 130 
and 190, respectively. Therefore, a sound theoretical description of 
the shell effects around N=50, 82 and 126 represents a major goal not 
only for nuclear structure physics but also for astrophysics.

%
%
\begin{figure*}
\includegraphics[width=1.0\textwidth]{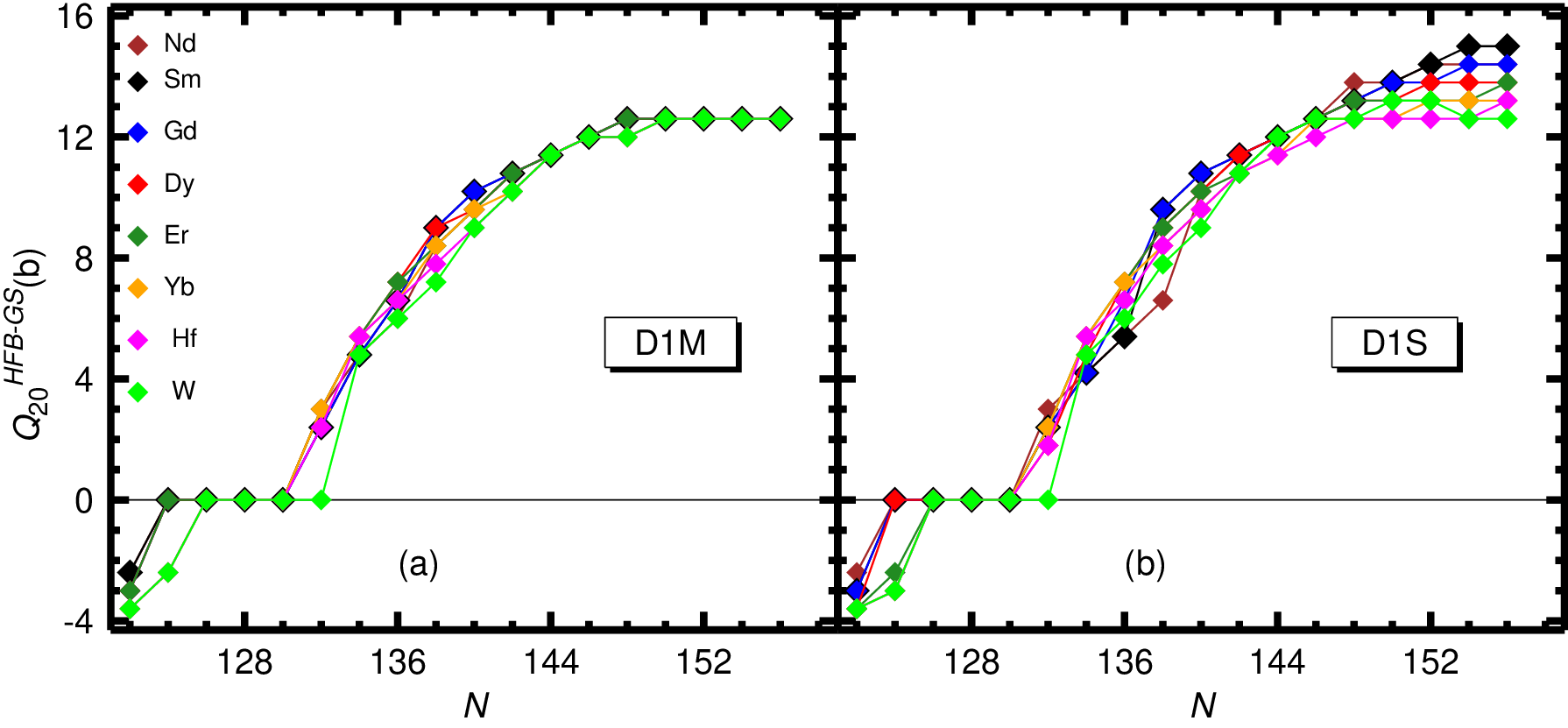}
\caption{ (Color online) The  ground 
state quadrupole deformations 
$Q_{20}^{HFB-GS}$
obtained for the  nuclei $^{182-216}$Nd, $^{184-218}$Sm, $^{186-220}$Gd, 
$^{188-222}$Dy, $^{190-224}$Er, $^{192-226}$Yb, $^{194-228}$Hf and $^{196-230}$W  
are plotted 
as a function of  neutron number. Results are shown for the Gogny-D1M [panel (a)]
and Gogny-D1S [panel (b)] EDFs. Calculations have been carried out with M$_{z,Max}$=14.
For more details, see the main text.
}
\label{fig-Q20-GS-summary}
\end{figure*}

It should be kept in mind that, due to their large neutron excess, very 
neutron-rich nuclei near the r-process path remain out of reach 
experimentally, especially those in the heavy mass region. Therefore, 
our understanding of those  systems requires the use of different 
theoretical tools whose predictions have been the subject of intense 
debate 
\cite{Doba-view1-r-proces,Doba-view2-r-proces,Sharma-view1-r-process,Sharma-view2-r-process,Sharma-view3-r-process}. 
For example, shell quenching has been invoked 
\cite{Pfeiffer-1997,Pearson-1996} at N=82 near the r-process path to 
improve the predicted r-process abundances around the second peak. The 
anomalous behavior of the 2$_{1}^{+}$ states in neutron-rich Cd 
isotopes has even  been cited \cite{DILLMANN} as an evidence for such a 
quenching. However, it has also been shown that this anomalous behavior 
can be naturally explained in the framework of symmetry-projected 
configuration mixing calculations \cite{2PLUS-CD} without the need to 
assume any  quenching of the N=82 shell closure. This is further 
corroborated by experimental results on energy levels in $^{130}$Cd 
\cite{Jungclaus-130Cd,Gorska-130Cd} .

In the present work, we have studied the quadrupole collectivity in the 
region of the third peak of the r-process. To this end, we have 
considered even-even Nd, Sm, Gd, Dy, Er, Yb, Hf and W nuclei with 
neutron numbers 122 $\le$ N $\le$ 156 extending beyond the two-neutron 
dripline.  The constrained Hartree-Fock-Bogoliubov (HFB) approximation 
\cite{rs} is used to obtain the mean field potential energy curves 
(MFPECs) corresponding to the  HFB energies as a function of the 
quadrupole moment. The MFPECs offer a valuable starting point to 
understand the evolution of the shell structure across N=126 as well as 
the emergence of quadrupole deformations in the considered isotopic 
chains. In particular, it will be shown that the onset of mean field 
(i.e., static) quadrupole deformation after crossing the spherical 
N=126 neutron shell closure leads to a pronounced enhancement of the 
two-neutron separation energies as compared with the ones resulting 
from the spherical HFB scheme. 

%
%
\begin{figure}
\includegraphics[width=0.475\textwidth]{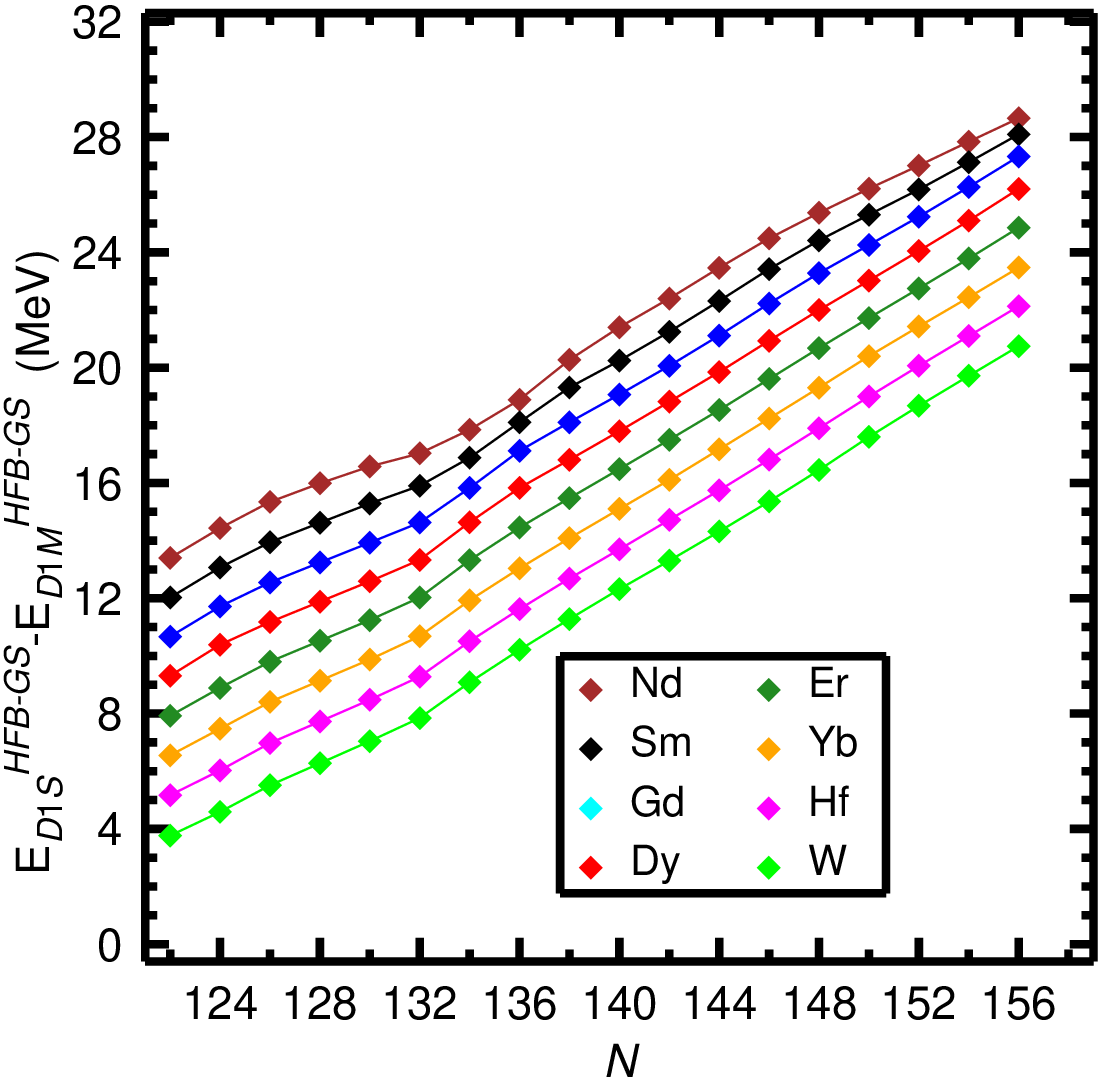}
\caption{ (Color online) The  differences 
E$_{D1S}^{HFB-GS}$-E$_{D1M}^{HFB-GS}$
between the HFB ground state energies obtained 
for the nuclei $^{182-216}$Nd, $^{184-218}$Sm, $^{186-220}$Gd, 
$^{188-222}$Dy, $^{190-224}$Er, $^{192-226}$Yb, $^{194-228}$Hf and $^{196-230}$W  
with the D1M and D1S parametrizations are plotted  as a function of 
neutron number. 
}
\label{fig-Diff-HFB-D1M-D1S} 
\end{figure}

The D1M parametrization \cite{gogny-d1m} of the Gogny  
\cite{Gogny-1980} EDF is used in all the HFB calculations. In some 
instances, however, results obtained with the  Gogny-D1S 
\cite{gogny-d1s} EDF will be shown for comparison. The reason is that 
the Gogny-D1S EDF has a strong reputation of being able to reproduce a 
large collection of low-energy nuclear data all over the nuclear chart 
both at the mean field level and beyond and related to deformation 
effects (see, for example, 
\cite{Rayner-Robledo-Egido-PRL,gogny-d1s,PRCQ2Q3-2012,PTpaper-Rayner,NPA-2002,Delaroche-2006,Bertsch-Peru-2007,Delaroche-Bertsch-2010} 
and references therein). However, it has to be kept in mind that D1S is 
not specially good in reproducing binding energies as shown in large 
scale calculations \cite{Hil-Gir-drift-2007,Martinez-Pinedo-D1SD1M} 
where a systematic drift is observed in the differences between the experimental 
and theoretical binding energies in heavy isotopes.

The D1M fitting protocol includes \cite{gogny-d1m} both realistic 
neutron matter equation of state (EoS) information and the binding 
energies of all known nuclei. With an impressive rms of 0.798 
MeV it represents a competitive choice  to deal with nuclear masses. In 
addition to previous studies \cite{gogny-d1m,PRCQ2Q3-2012,PTpaper-Rayner}, 
new ones including fission properties in heavy and superheavy nuclei 
\cite{fission-U-Rayner-Robledo,fission-Pu-Rayner-Robledo} as well as 
odd nuclei within the equal filling approximation (EFA) 
\cite{Sara-odd,odd-D1M-ref1,odd-D1M-ref2}, suggest that the Gogny-D1M 
EDF  essentially retains the  predictive power of the well tested D1S 
parametrization and therefore it represents a good candidate to replace the 
latter.

The MFPECs obtained within the Gogny-HFB framework display, at least 
for some Nd, Sm, Gd, Dy, Er, Yb, Hf and W nuclei, competing minima 
based on different intrinsic configurations indicating that beyond 
mean field correlations like symmetry restoration and/or quadrupole 
configuration mixing, may play a role. That the restoration of the 
broken rotational symmetry may provide leading quantum corrections can 
be expected from the fact that the energy gain associated with it 
(i.e., the so called rotational correction) is proportional to the 
deformation of the intrinsic HFB states 
\cite{PLB-Rayner-Egido-Robledo,Valor-24Mg}. Similar to what has been 
found in other regions of the nuclear chart 
\cite{NPA-2002,PLB-Rayner-Egido-Robledo,RaynerN20-PRC2000,RaynerN28}, 
symmetry restoration also leads in some of the studied  nuclei to 
important topological changes in the angular momentum projected 
potential energy curves (AMPPECs) as compared with the corresponding 
MFPECs. Therefore, it is  also important to consider the effect of the 
quadrupole configuration mixing in those cases. Moreover, even in those 
nuclei where the AMPPECs exhibit well defined minima, it is important 
to check their stability with respect to quadrupole fluctuations since 
not only the energy landscape but also the underlying collective 
inertia play a role within a dynamical treatment. With this in mind we 
have performed angular momentum projected generator coordinate method 
(AMPGCM) calculations \cite{NPA-2002}, based on the Gogny-D1M EDF, for 
all the nuclei studied in this work. In addition to the spectroscopic 
properties of the  excited state  the AMPGCM also provides the ground 
state correlation energy that, as we will see, plays an important role in the properties 
of the N=126 shell closure.

The paper is organized as follows. In Sec. \ref{Theory}, we describe 
the theoretical approximations used. The key ingredients of our HFB 
approach are presented in Sec. \ref{HFB-theory} while the AMPGCM 
formalism \cite{NPA-2002} is briefly outlined in Sec. 
\ref{AMPGCM-theory}. The  results of the calculations are discussed in 
Sec. \ref{RESULTS}. First, in Sec. \ref{RESULTS-HFB}, we discuss the 
static quadrupole properties predicted within our HFB calculations. The 
results of our symmetry-projected configuration mixing study are 
presented in Sec. \ref{RESULTS-BYMF}. Finally, Sec. \ref{conclusions} 
is devoted to the concluding remarks and work perspectives.


\section{Theoretical framework}
\label{Theory}

In this section, the theoretical approximations used in the present 
work are described. First, in Sec. \ref{HFB-theory}, the HFB framework 
\cite{rs} is introduced. Next the AMPGCM formalism \cite{NPA-2002} is 
briefly outlined in Sec. \ref{AMPGCM-theory}. For a more detailed 
description of both angular momentum projection (AMP) and the Generator 
Coordinate Method (GCM) the reader is referred to the literature 
\cite{PLB-Rayner-Egido-Robledo,Bonche-mixed,Schmid-1,Schmid-2,Hara-lambda,Hara-Sun-AMP}.

%
%
\begin{figure}
\includegraphics[width=0.475\textwidth]{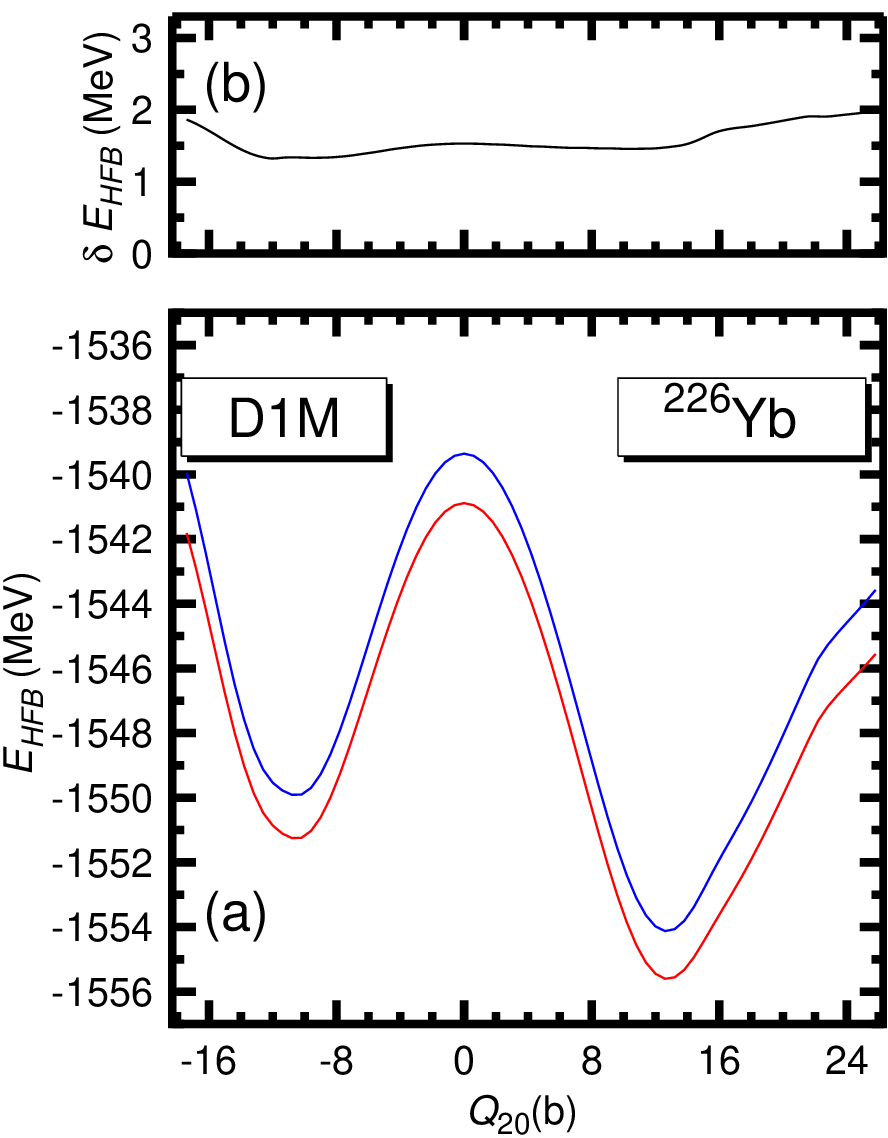}
\caption{ (Color online) In panel (a) the HFB energies computed with b$_{0}$=2.1 fm and M$_{z,Max}$=14 (blue)
and M$_{z,Max}$=17 (red) are plotted as a function of the quadrupole moment.
Results are shown for the nucleus $^{226}$Yb using the D1M parametrization. In panel (b), the energy difference 
between the M$_{z,Max}$=14 and M$_{z,Max}$=17 calculations is plotted as a function of the quadrupole 
moment.
}
\label{fig-14vs17-D1M-226Yb} 
\end{figure}

\subsection{The Hartree-Fock-Bogoliubov approximation}
\label{HFB-theory}

The starting point is the (constrained) HFB approximation 
\cite{rs,Rayner-Robledo-Egido-PRL} for the finite range and density 
dependent Gogny-EDF \cite{Gogny-1980}. Both the D1S \cite{gogny-d1s} 
and D1M \cite{gogny-d1m} parametrizations are considered. As 
constraining operators the axially symmetric quadrupole 
$\hat{Q}_{20}=z^{2}-\frac{1}{2}\left(x^{2}+y^{2} \right)$ 
\cite{PRCQ2Q3-2012,PTpaper-Rayner} moment as well as the standard HFB 
constraints on both the proton $\hat{Z}$ and neutron $\hat{N}$ number 
operators are used. The HFB quasiparticle operators are expanded in an 
axially symmetric and parity-preserving harmonic oscillator  (HO) basis 
whose quantum numbers are restricted by the condition

\begin{eqnarray}
2n_{\perp} + |m| +\frac{1}{q} n_{z} \le M_{z,Max}
\end{eqnarray}

All the mean field results to be discussed later on have been obtained 
with q=1.0. The two length parameters $b_{z}$ and $b_{\perp}=b_{0}$ 
characterizing the HO basis are chosen to be equal to keep the basis 
closed under rotations \cite{NPA-2002,WickRobledo} (this is also the 
reason to include full HO major shells in the basis). This simplifies 
the application of the AMPGCM approximation described in the next Sec. 
\ref{AMPGCM-theory}. Most of our results have been obtained with 
M$_{z,Max}$=14. However, we have also tested the stability of our 
predictions with respect to the size of the considered single-particle 
basis by performing  HFB calculations with M$_{z,Max}$=17.  

%
%
\begin{figure*}
\includegraphics[width=0.4850\textwidth]{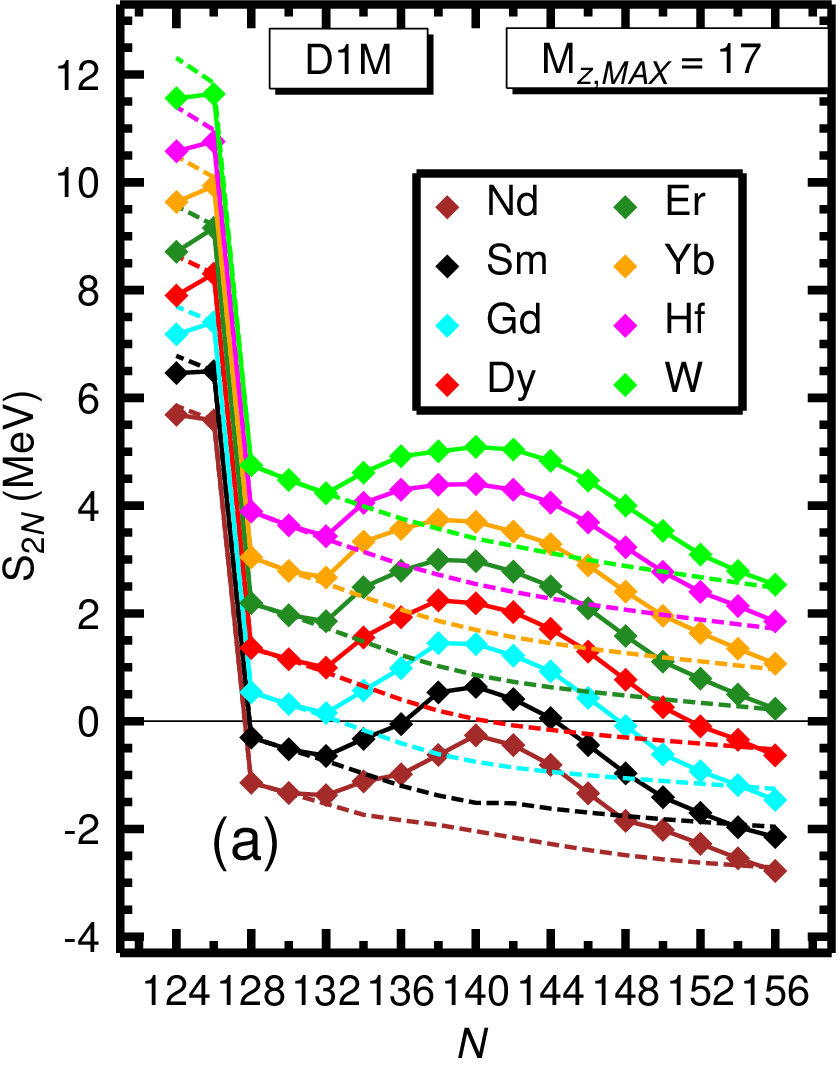}
\includegraphics[width=0.4560\textwidth]{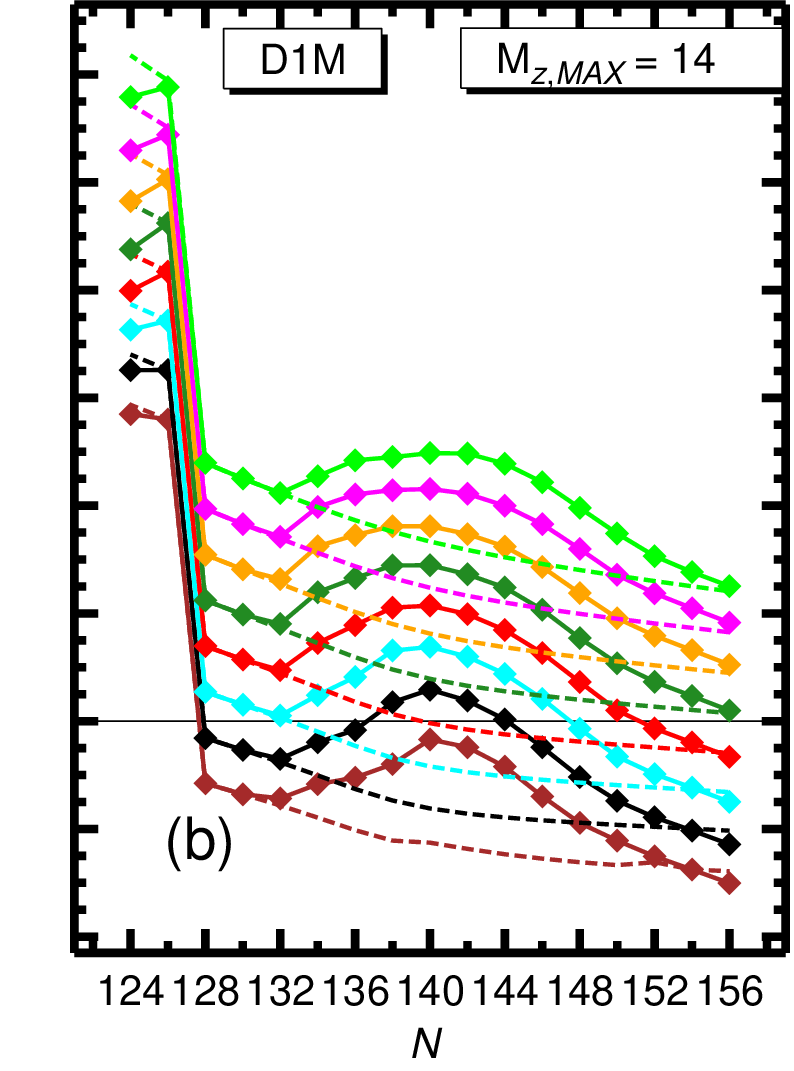} 
\caption{ (Color online) The two-neutron separation energies (full lines) computed within the HFB 
approximation for the nuclei 
$^{182-216}$Nd, $^{184-218}$Sm, $^{186-220}$Gd, $^{188-222}$Dy, $^{190-224}$Er, $^{192-226}$Yb,
$^{194-228}$Hf and $^{196-230}$W are plotted as a function of the neutron number. 
Results based on the Gogny-D1M EDF and the bases   M$_{z,Max}$=17 
and  M$_{z,Max}$=14 are shown in panels (a) and (b), respectively. The two-neutron separation 
energies obtained in the framework of spherical HFB calculations are also 
included in the plots (dashed lines) for comparison.
For more details, see the main text.
}
\label{fig-S2N-shells-D1M} 
\end{figure*}

It should be kept in mind that for very neutron-rich nuclei, especially 
those in the proximity of the two-neutron dripline, the HFB 
approximation must be used \cite{Doba-dripline-1996,Doba-dripline-1994} 
and absolute convergence for the binding energy can only be obtained 
for HO bases with a very large number of shells  M$_{z,Max}$. At the 
HFB level such a computationally demanding task can be afforded with 
present day computer capabilities. Also in the case that we were just 
interested in a single $Q_{20}$-configuration, AMP calculations with a 
very large M$_{z,Max}$ value could be afforded. However, the 
considerable amount of angular momentum projected Hamiltonian kernels 
to be computed in the AMPGCM calculations (see, Sec. 
\ref{AMPGCM-theory}), restrict the maximum M$_{z,Max}$ value to 
M$_{z,Max}=14$. The reason behind is the finite range component of the 
Gogny-EDF  that makes the evaluation of the corresponding matrix 
elements very time consuming. On the other hand, as the collective 
motion is mainly affected by the shape of the energy landscape, not its 
absolute depth, a quicker convergence with M$_{z,Max}$ is achieved for 
the related physical quantities.

The   HFB equations have been solved with the help of an approximate 
second order gradient method  \cite{Robledo-Bertsch2OGM} which allows 
us to handle  constraints efficiently 
\cite{PRCQ2Q3-2012,PTpaper-Rayner,Robledo-Rayner-JPG-2012,fission-U-Rayner-Robledo,fission-Pu-Rayner-Robledo}. 
The method is based on the parametrization of a given HFB vacuum in 
terms of the Thouless theorem \cite{rs}. Recently, similar variational 
strategies have been  applied to  correlated electronic systems in 
condensed matter physics 
\cite{rayner-Hubbard-1D-FED2013,rayner-Hubbard-1D-FED2014} and quantum 
chemistry 
\cite{Carlos-Rayner-Gustavo-VAMPIR-molecules,Carlos-Rayner-Gustavo-FED-molecules}.

%
%
\begin{figure}
\includegraphics[width=0.4850\textwidth]{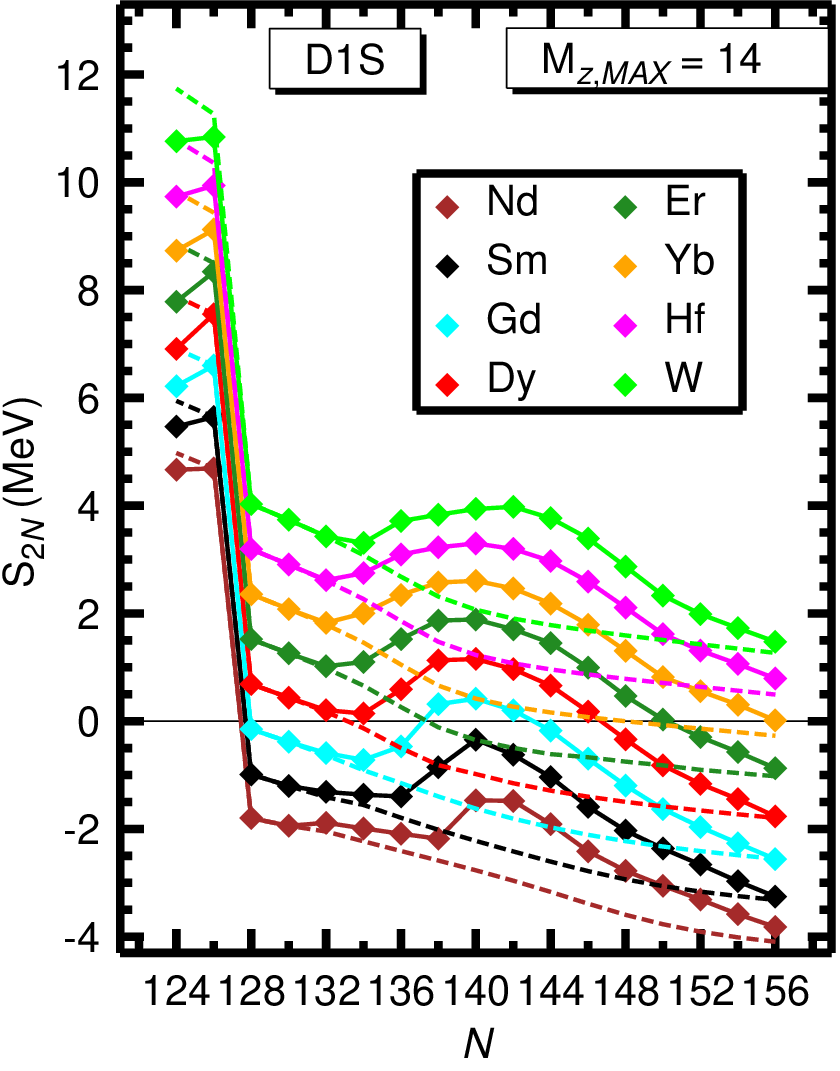}
\caption{ (Color online) The two-neutron separation energies (full lines) computed within the 
Gogny-D1S HFB approximation for the nuclei 
$^{182-216}$Nd, $^{184-218}$Sm, $^{186-220}$Gd, $^{188-222}$Dy, $^{190-224}$Er, $^{192-226}$Yb,
$^{194-228}$Hf and $^{196-230}$W are plotted as a function of neutron number.   
Results have been obtained with M$_{z,Max}$=14. The two-neutron separation 
energies obtained in the framework of spherical HFB calculations are also 
included in the plots (dashed lines) for comparison.
}
\label{fig-S2N-shells-D1S} 
\end{figure}

The constrained HFB approximation provides us  with a set of mean field 
product states $|\varphi (Q_{20}) \rangle$, labeled by the generating 
coordinate $\langle \varphi | \hat{Q}_{20} |  \varphi \rangle = 
Q_{20}$, as well as the MFPECs. In our  calculations we have used the 
grid -26 b $\le$ $Q_{20}$ $\le$ 36 b with a mesh size  $\delta$ 
$Q_{20}$ = 0.6 b. We have tested that they are accurate enough to 
describe the low-energy quadrupole dynamics of the nuclei considered. 
In particular, as will be shown, the selected $Q_{20}$-grid is  enough 
for the AMPGCM collective wave functions to reach the zero value in 
their tails (see, Sec.\ref{RESULTS-BYMF}).

Other interesting pieces of information coming from the mean field 
approximation are the proton and neutron single-particle energies 
(SPEs). The quadrupole deformation effects are strongly linked to the 
position of the Fermi energies in such spectra  
\cite{PTpaper-Rayner,Butler-Naza,Valdo-Robledo-spart,Yo-JPG-2010}. We 
have studied the evolution of the proton and neutron SPEs with the 
quadrupole moment $Q_{20}$. To this end, we have diagonalized the 
Routhian $h=t + \Gamma -\lambda_{Q_{20}} Q_{20}$, with $t$ being the 
kinetic energy operator and $\Gamma$ the Hartree-Fock field \cite{rs}. 
The term $\lambda_{Q_{20}} Q_{20}$ contains the Lagrange multiplier 
used to enforce the corresponding quadrupole constraint.

%
%
\begin{figure*}
\includegraphics[width=0.5\textwidth,angle=-90]{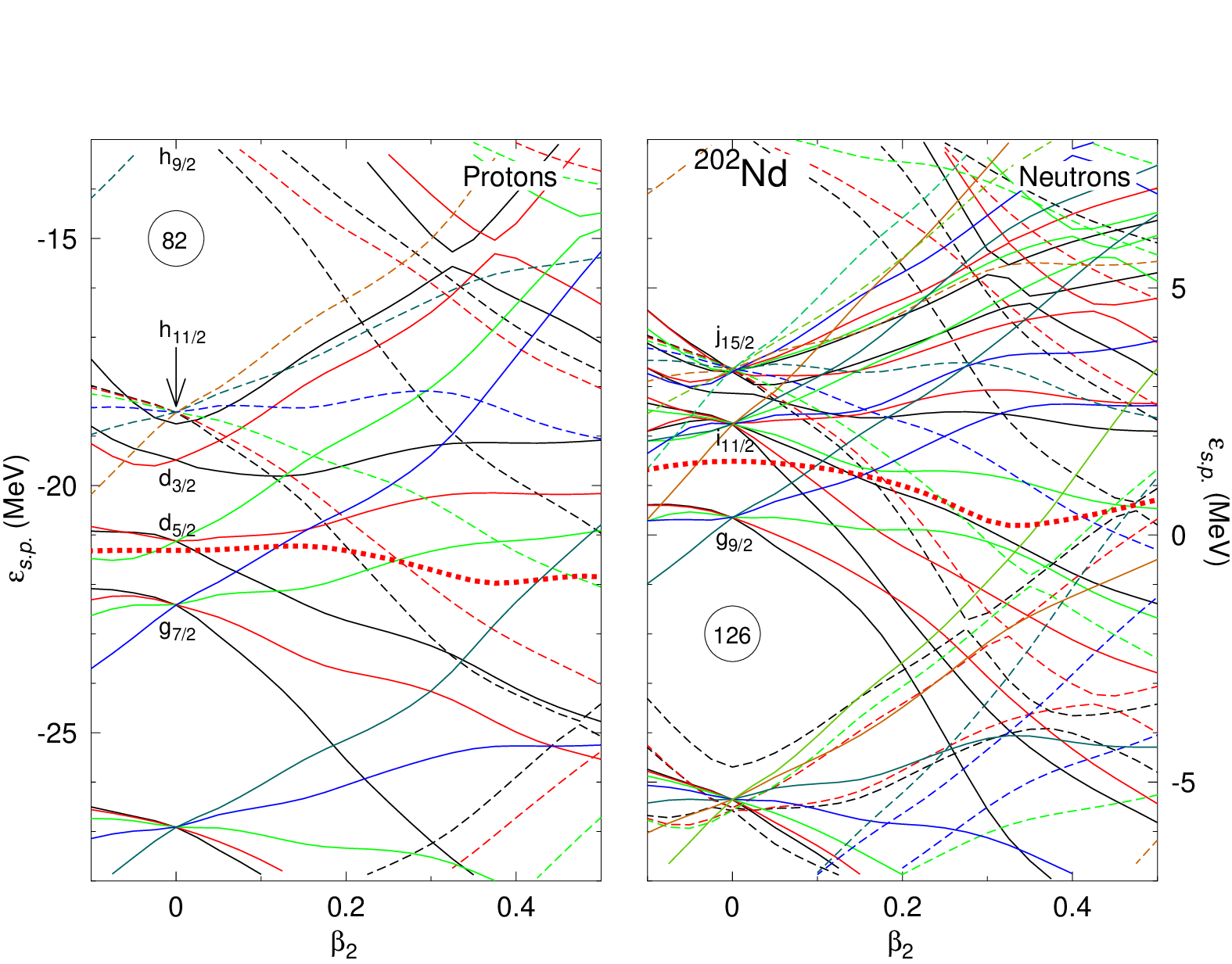}
\includegraphics[width=0.5\textwidth,angle=-90]{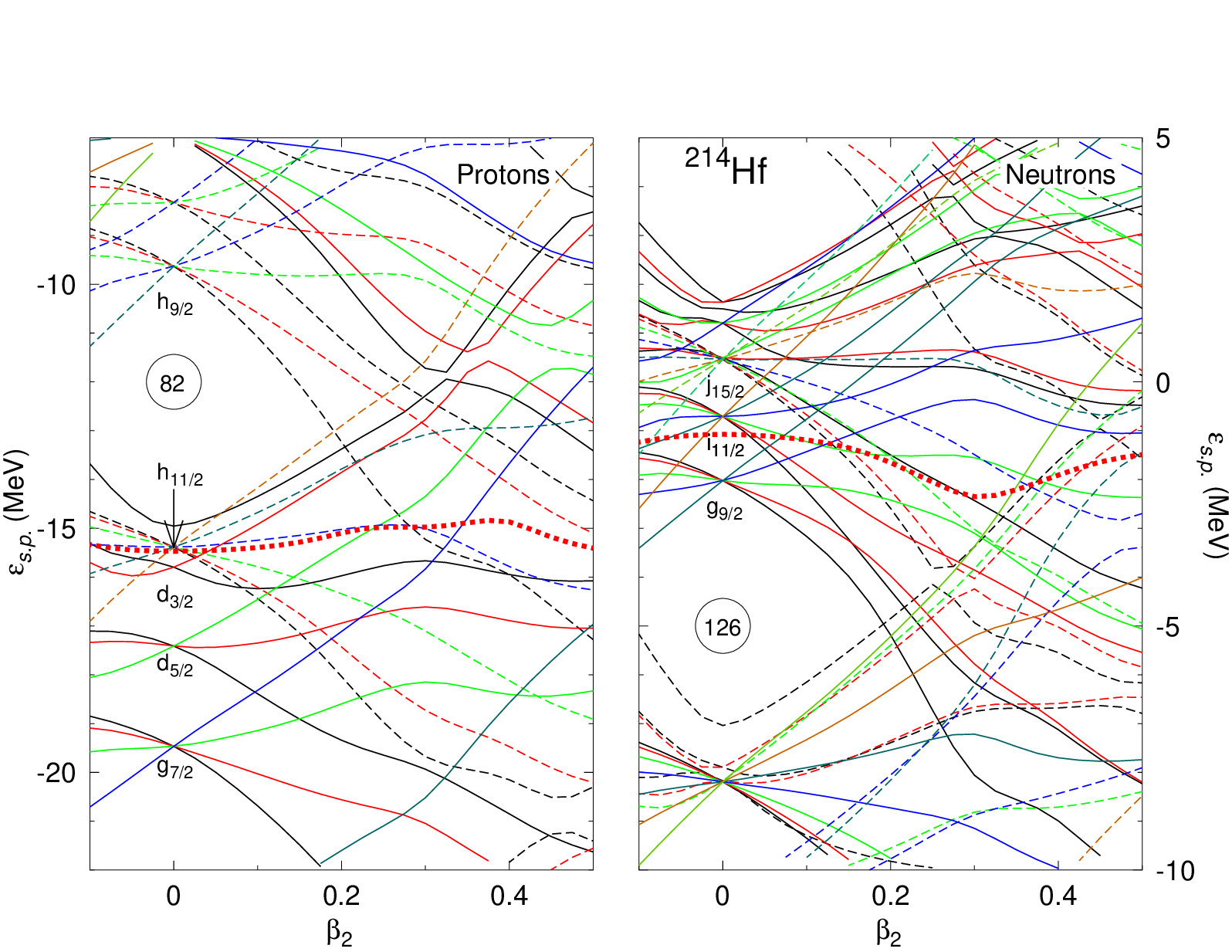}
\caption{ (Color online) The proton and neutron SPEs, computed with the Gogny-D1M 
 EDF, are plotted as a function of the $\beta_{2}$ deformation parameter for the N=142 nuclei $^{202}$Nd and $^{214}$Hf.
Full (dashed) lines correspond to positive (negative) parity orbitals. Different $K$ projections
have associated a different color (black, red, green, blue, dark-green, brown, etc for $K=1/2, 3/2, \ldots$. The
Fermi level is plotted with a thick red dashed line.}
\label{fig-SPE} 
\end{figure*}

\subsection{Symmetry-projected quadrupole configuration mixing}
\label{AMPGCM-theory}

Having a set of symmetry breaking (i.e., intrinsic) HFB states 
$|\varphi (Q_{20}) \rangle$ at hand, one introduces the following 
AMPGCM ansatz

\begin{eqnarray} \label{GCM-state}
| \Psi_{M,\sigma}^{I} \rangle = \sum_{K} \int d Q_{20} f_{K; \sigma}^{I}(Q_{20}) \hat{P}_{MK}^{I} |\varphi (Q_{20}) \rangle
\end{eqnarray}
which superposes the  symmetry-projected states $\hat{P}_{MK}^{I} |\varphi (Q_{20}) \rangle$ 
with amplitudes 
$f_{K;\sigma}^{I}(Q_{20})$. The projection operator reads
\cite{rs}

\begin{eqnarray}
\hat{P}_{MK}^{I} = \frac{2I+1}{8 \pi^{2}} \int d \Omega D_{MK}^{I *} \hat{R}(\Omega)
\end{eqnarray}
where $ R(\Omega)= e^{-i \alpha \hat{J}_{z}} e^{-i \beta \hat{J}_{y}} 
e^{-i \gamma \hat{J}_{z}}$ is the rotation operator, $\Omega = 
\left(\alpha, \beta, \gamma \right)$ stands for the set of Euler angles 
and ${\cal{D}}_{M K}^{I}(\Omega)$ are Wigner functions  \cite{Edmonds}. 
Our set of generating states $|\varphi (Q_{20}) \rangle$ only 
comprises axially symmetric and parity-preserving K=0 HFB vacua 
(quasiparticle excitations \cite{Hara-Sun-AMP} are not included). 
As a result, the integrals over $\alpha$ and $\gamma$ can be carried 
out analytically and one is only left with the numerical 
$\beta$-integration over a suitable grid. The amplitudes 
$f_{K=0;\sigma}^{I}(Q_{20})= f_{\sigma}^{I}(Q_{20})$ are then 
determined through the solution of the Hill-Wheeler (HW) equation 
\cite{NPA-2002,HW-ref}

\begin{small}
\begin{eqnarray} \label{HW}
\int d Q_{20}^{'} \Big[{\cal{H}}^{I}(Q_{20},Q_{20}^{'}) -E_{\sigma}^{I} 
{\cal{N}}^{I}(Q_{20},Q_{20}^{'})\Big] f_{\sigma}^{I}(Q_{20}^{'}) =0
\end{eqnarray}
\end{small}
where 

\begin{small}
\begin{eqnarray} \label{Kernels}
{\cal{H}}^{I}(Q_{20},Q_{20}^{'}) &=& \Delta(I) 
\left( 2I +1 \right) \int_{0}^{\pi/2} d\beta \sin(\beta) h(\beta)
\nonumber\\
{\cal{N}}^{I}(Q_{20},Q_{20}^{'}) &=& \Delta(I)
\left( 2I +1 \right) \int_{0}^{\pi/2} d\beta \sin(\beta) n(\beta)
\nonumber\\
\Delta(I) &=& \frac{1}{2} \left(1+ (-)^{I} \right)
\nonumber\\
h(\beta) &=&
\langle \varphi (Q_{20}) | \hat{H} \Big[\rho_{\beta}^{Mix}(\vec{r}) \Big] e^{-i \beta \hat{J}_{y}} | \varphi (Q_{20}^{'}) \rangle \nonumber\\
n(\beta) &=&
\langle \varphi (Q_{20}) | e^{-i \beta \hat{J}_{y}} | \varphi (Q_{20}^{'}) \rangle
\end{eqnarray}
\end{small}
and $\rho_{\beta}^{Mix}(\vec{r})$ is the so-called {\it{mixed density}} prescription, i.e.,

\begin{eqnarray}
\rho_{\beta}^{Mix}(\vec{r}) = 
\frac{\langle \varphi(Q_{20}) | \hat{\rho}(\vec{r}) e^{-i \beta \hat{J}_{y}} |  \varphi(Q_{20}^{'} \rangle}{\langle \varphi(Q_{20}) 
|  e^{-i \beta \hat{J}_{y}} |
\varphi(Q_{20}^{'}) \rangle}
\end{eqnarray}
widely  used in the context of symmetry restoration and/or configuration mixing 
\cite{PRCQ2Q3-2012,NPA-2002,Bender-review,Bonche-mixed,Bender-other-1,Duguet,Rayner-Pb,Robledo-dens}. 
Since the average values of the proton and neutron numbers usually differ 
from the nucleus' proton Z$_{0}$ and neutron  N$_{0}$ numbers we have 
replaced $\hat{H}$ by 
$\hat{H}-\lambda_{Z} \left(\hat{Z} - Z_{0}\right)-\lambda_{N} \left(\hat{N} - N_{0}\right)$, 
where $\lambda_{Z}$ and $\lambda_{N}$ are chemical potentials for protons 
and neutrons, respectively \cite{Bonche-mixed,Egido-lambda,Hara-lambda}. 
All the  AMPGCM calculations discussed in this work have been performed 
with the Gogny-D1M EDF and M$_{z,Max}$=14.

%
%
\begin{figure*}
\includegraphics[width=1.0\textwidth]{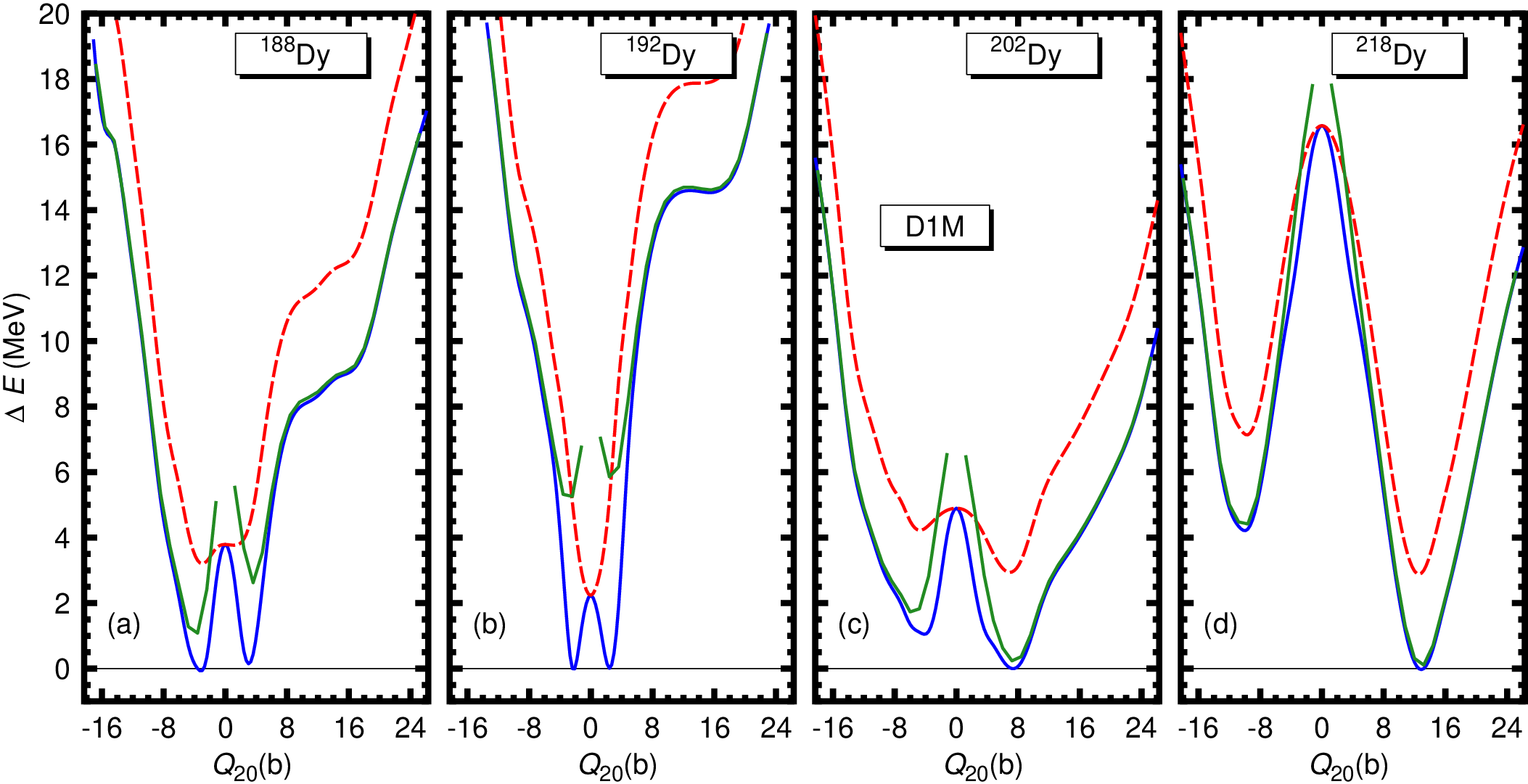}
\caption{
(Color online) The AMPPECs   for the nuclei $^{188,192,202,218}$Dy and 
the spin values $I^{\pi}=0^{+}$ (blue) and  $2^{+}$ (green) are 
depicted as a function of the quadrupole moment $Q_{20}$. The MFPECs 
(dashed red) are also included in the plots for comparison. All the 
energies are referred to the absolute minimum of the $I^{\pi}=0^{+}$ 
AMPPECs. Results have been obtained with the parametrization D1M of the 
Gogny-EDF.
}
\label{fig-AMPPECs-Dy} 
\end{figure*}

For a given spin $I$, the solution of the HW equation (\ref{HW}) provides the energies E$_{\sigma}^{I}$ corresponding to the 
ground ($\sigma$=1) and excited ($\sigma$=2,3 $\dots$) states. However, since the
symmetry-projected basis states used in the expansion Eq.(\ref{GCM-state})
are not orthogonal, the functions $f_{\sigma}^{I}(Q_{20})$
cannot be interpreted as probability amplitudes. One then introduces \cite{PRCQ2Q3-2012,NPA-2002} 
the collective wave functions 

\begin{eqnarray} \label{COLLWFS}
G_{\sigma}^{I}(Q_{20}) = \int d Q_{20}^{'} {\cal{N}}^{I \frac{1}{2}}(Q_{20},Q_{20}^{'})  f_{\sigma}^{I}(Q_{20}^{'})
\end{eqnarray}
in terms of the operational square root of the norm kernel
\cite{NPA-2002,rs}. The collective wave functions 
Eq. (\ref{COLLWFS}) are orthogonal and their modulus squared $|G_{\sigma}^{I}(Q_{20})|^{2}$
has the meaning of a probability amplitude \cite{rs}. In order
to understand these collective wave functions in a more quantitative 
way, we have computed the averages \cite{NPA-2002}

\begin{eqnarray} \label{AMPGCM-ave-Q20}
Q_{20}^{I, \sigma} =
\int dQ_{20} |G_{\sigma}^{I}(Q_{20})|^{2} Q_{20}
\end{eqnarray}
providing us with a measure of the deformation in the underlying intrinsic
states.


\section{Discussion of results}
\label{RESULTS}

In this section, the results of our calculations are discussed. First, in 
Sec. \ref{RESULTS-HFB}, the static quadrupole properties 
obtained within our Gogny-HFB framework are addressed. The results of the 
AMPGCM calculations with Gogny-D1M  are presented in Sec. \ref{RESULTS-BYMF}.

\subsection{The HFB approximation: static quadrupole properties}
\label{RESULTS-HFB}

In Fig. \ref{fig-PEC-Sm_D1M}, we have plotted [panels (a)-(r)] a 
typical outcome of the constrained HFB calculations based on the Gogny 
D1M and D1S EDFs.  The MFPECs are shown for the nuclei $^{184-218}$Sm 
taken as illustrative examples. For the sake of clarity they are 
depicted for -18.3 b $\le$ $Q_{20}$ $\le$ 26.3 b. The HFB energies are 
always referred to the absolute minimum of the corresponding  MFPEC. 
Calculations have been carried out with M$_{z,Max}$=14. 

It is apparent from the figure that, though the energies $\Delta 
E_{HFB}$ tend to be smaller with the parametrization D1S, both 
Gogny-EDFs provide MFPECs with a similar structure. The nucleus 
$^{184}$Sm displays a slightly oblate  ground state 
($Q_{20}^{HFB-GS}$=-2.4 b and $Q_{20}^{HFB-GS}$=-3.0 b with the D1M and 
D1S parametrizations, respectively) while spherical ground states are 
predicted for  $^{186-192}$Sm. In the case of $^{188}$Sm, the MFPEC 
exhibits the clear signature of a strong N=126 neutron shell closure 
with vanishing neutron pairing energy at $Q_{20}$=0. Already for the 
isotope $^{194}$Sm (i.e., N=132), the HFB ground state becomes slightly 
prolate deformed with $Q_{20}^{HFB-GS}$=2.4 b. Prolate deformed ground 
states are also predicted for the heavier isotopes $^{196-218}$Sm.  
Typical $\beta_{2}$ values for this chain are $\beta_{2}$=-0.07 for 
$^{184}$Sm,  $\beta_{2}$= 0.06 for $^{194}$Sm and $\beta_{2}$= 0.28 for 
$^{214}$Sm, all of them with the parametrization D1M. Similar results are found for 
the nuclei $^{182-216}$Nd, $^{186-220}$Gd, $^{188-222}$Dy, 
$^{190-224}$Er, $^{192-226}$Yb, $^{194-228}$Hf and $^{196-230}$W as can 
be seen from Fig. \ref{fig-Q20-GS-summary} where the ground state 
deformations $Q_{20}^{HFB-GS}$ have been plotted as a function of 
neutron number for the Gogny-D1M [panel (a)] and Gogny-D1S [panel (b)] 
EDFs.  

%
%
\begin{figure*}
\includegraphics[width=0.75\textwidth]{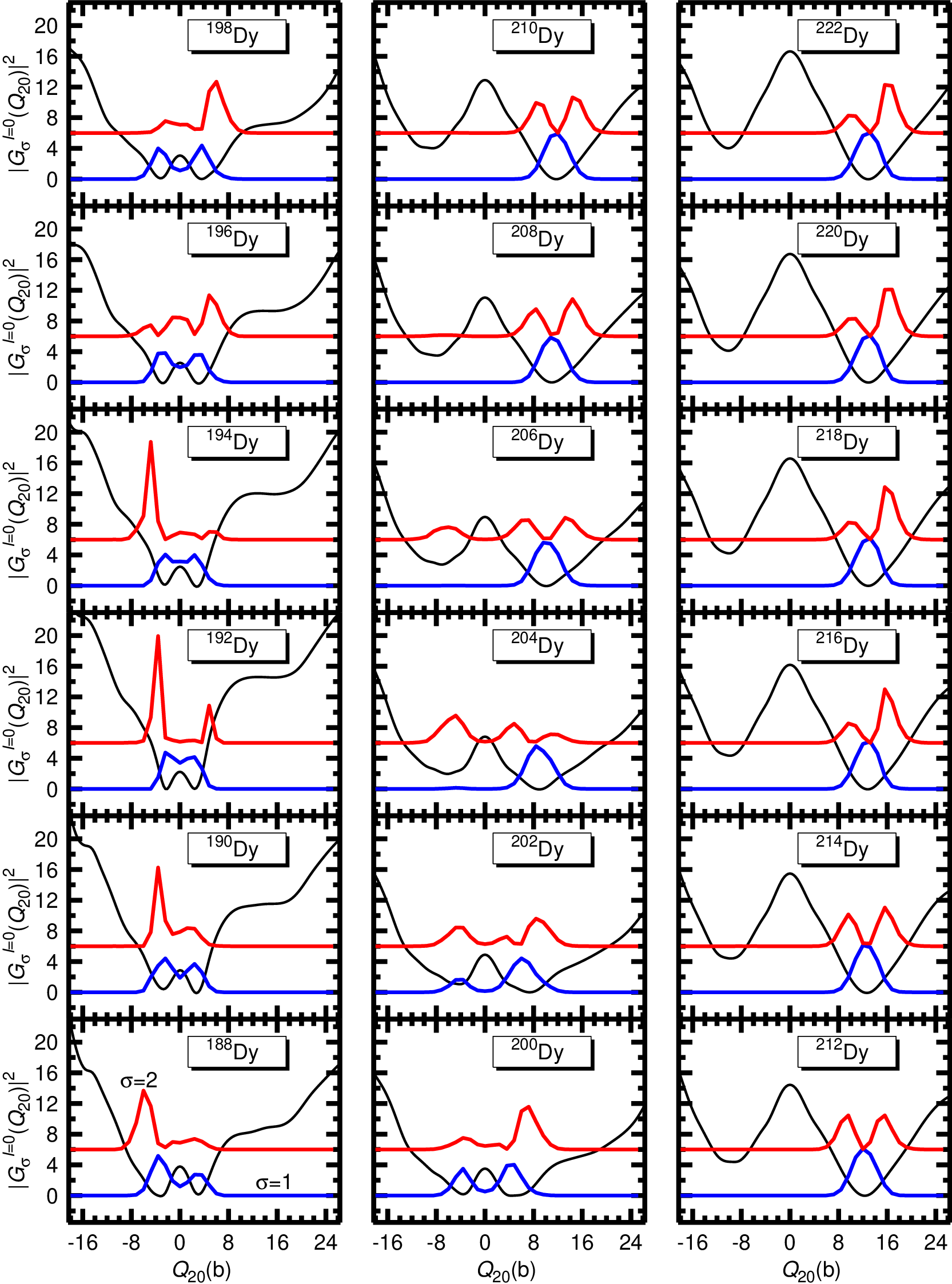}
\caption{ (Color online) The square of the I$^{\pi}$=0$^{+}$ collective 
wave functions [Eq.(\ref{COLLWFS})] corresponding to the ground 
($\sigma$=1) and first excited ($\sigma$=2) states in the nuclei 
$^{188-222}$Dy are plotted as a function of the quadrupole moment 
$Q_{20}$.  The quantities actually depicted are 
25 $\times$ $|$ G$_{\sigma=1}^{I=0}$($Q_{20}$)$|^{2}$
(blue) and 6 + 25 $\times$ $|$ G$_{\sigma=2}^{I=0}$($Q_{20}$)$|^{2}$ (red). 
For each nucleus the I$^{\pi}$=0$^{+}$ AMPPECs (black)
are also included in the plots. Energies are always referred to the 
absolute minima of the corresponding AMPPECs. Results have been obtained 
with the parametrization D1M of the Gogny-EDF.
}
\label{fig-WFs-Dy} 
\end{figure*}

Therefore, in the framework of the Gogny-HFB calculations, N=126 
appears as a spherical neutron magic number. However,  6-8 
mass units beyond N=126, regardless of the Gogny-EDF employed, there is 
an onset of prolate deformation in the ground states along all the studied 
isotopic chains.  

This effect agrees well with the additional stability beyond N=126 
predicted in the relativistic mean field theory (RMF) based on the 
NL-SV1 Lagrangian model with the inclusion of the vector self-coupling 
of $\omega$-meson 
\cite{N126-Farhan-2006,N126-Sharma2011,FarhanSharma-tobe-published}. A 
similar effect can also be seen in the mass formulas HFB-14 based upon 
Skyrme-HFB \cite{Goriely-HFB-14} and in the finite-range droplet model 
(FRDM) \cite{FRDM-mass-1994}.

%
%
\begin{figure*}
\includegraphics[width=0.95\textwidth]{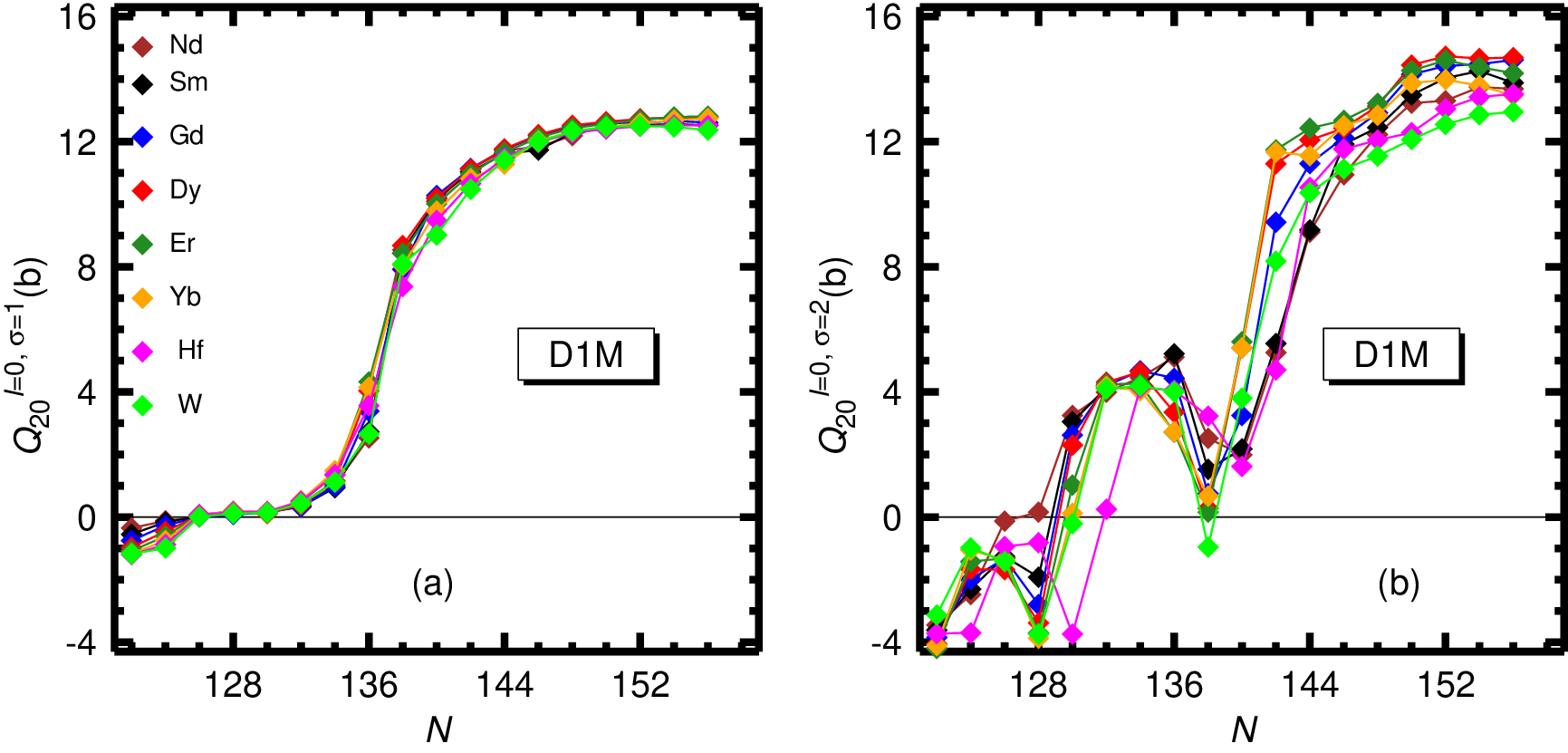}
\caption{ 
(Color online) The intrinsic   deformations [Eq.(\ref{AMPGCM-ave-Q20})]
associated with the ground ($\sigma$=1) and first excited  
($\sigma$=2) states  at I$^{\pi}$=0$^{+}$ in the nuclei
$^{182-216}$Nd, $^{184-218}$Sm, $^{186-220}$Gd, $^{188-222}$Dy, 
$^{190-224}$Er, $^{192-226}$Yb, $^{194-228}$Hf and $^{196-230}$W 
are depicted in panels (a) and (b), respectively, as a function of neutron number.
Results have been obtained with the parametrization D1M of the Gogny-EDF.
For more details, see the main text.
}
\label{fig-QAMPGCM-gs-summary} 
\end{figure*}

From Figs. \ref{fig-PEC-Sm_D1M} and \ref{fig-Q20-GS-summary} we 
conclude that both the D1M and D1S parametrizations provide quite 
similar deformation effects for the considered nuclei. However, the D1S 
results show a pronounced under-binding as can be seen in Fig. 
\ref{fig-Diff-HFB-D1M-D1S} where the ground state energy differences 
E$_{D1S}^{HFB-GS}$-E$_{D1M}^{HFB-GS}$ are plotted as a function of 
neutron number. This quantity increases almost linearly with neutron 
number while it decreases for increasing Z values. These results are 
not surprising as they reflect a well known deficiency 
\cite{Hil-Gir-drift-2007} of the Gogny-D1S EDF away from the stability 
valley. A pronounced under-binding has also been found in the Gogny-D1S 
fission paths, as compared with the D1M ones, obtained in our recent 
studies  of neutron-rich U and Pu nuclei as well as in superheavy 
elements \cite{fission-U-Rayner-Robledo,fission-Pu-Rayner-Robledo}. The 
linear increase of the D1S under-binding with neutron number implies 
that the behavior of the two-neutron separation energies will be 
similar to that with D1M but the quantity will be shifted down  by 
roughly 1 MeV.

A few words  concerning the convergence of our calculations are in 
order here. In panel (a) of  Fig. \ref{fig-14vs17-D1M-226Yb}, we have 
plotted the HFB energies, computed with  M$_{z,Max}$=14  and 
M$_{z,Max}$=17, as a function of the quadrupole moment. In panel 
(b), we have plotted the energy difference betweeen both calculations. 
Results are shown for $^{226}$Yb and the Gogny-D1M EDF but similar ones 
are obtained for other nuclei and/or the D1S parametrization. In the 
range -18.3 b $\le$ $Q_{20}$ $\le$ 26.3 b the energy landscape does not 
change much when the size of the basis is increased. As we will see in 
Sec. \ref{RESULTS-BYMF}, this range of quadrupole deformations is the 
one where the collective dynamics  concentrates and, therefore, no 
significant differences are expected between the AMPGCM calculations 
with M$_{z,Max}$=14 and M$_{z,Max}$=17. Since the main interest of the 
present study is focussed on the two-neutron separation energies 
S$_{2N}$, rotational energy corrections, excitation energies, etc., and 
these quantities do not change very much with the considered   
M$_{z,Max}$ value, we conclude that M$_{z,Max}$=14 can be regarded as a 
reasonable compromise between accuracy and computational burden. To corroborate 
this conclusion, the two-neutron separation energies (full lines) computed within the 
HFB approximation are plotted in Fig. \ref{fig-S2N-shells-D1M} as a 
function of  neutron number. Results based on the Gogny-D1M EDF and the 
M$_{z,Max}$=17 and  M$_{z,Max}$=14 bases are shown in panels (a) and 
(b), respectively. The two-neutron separation energies obtained by 
restricting to spherical HFB calculations are also included in the 
plots (dashed lines) for comparison. One sees that the trends and 
numerical values predicted within the M$_{z,Max}$=17 and M$_{z,Max}$=14 
calculations are quite similar, which corroborates  our choice of 
M$_{z,Max}$=14  as a reasonable basis size to study the physical 
properties we are interested in this work.

\begin{figure*}
\includegraphics[width=0.75\textwidth]{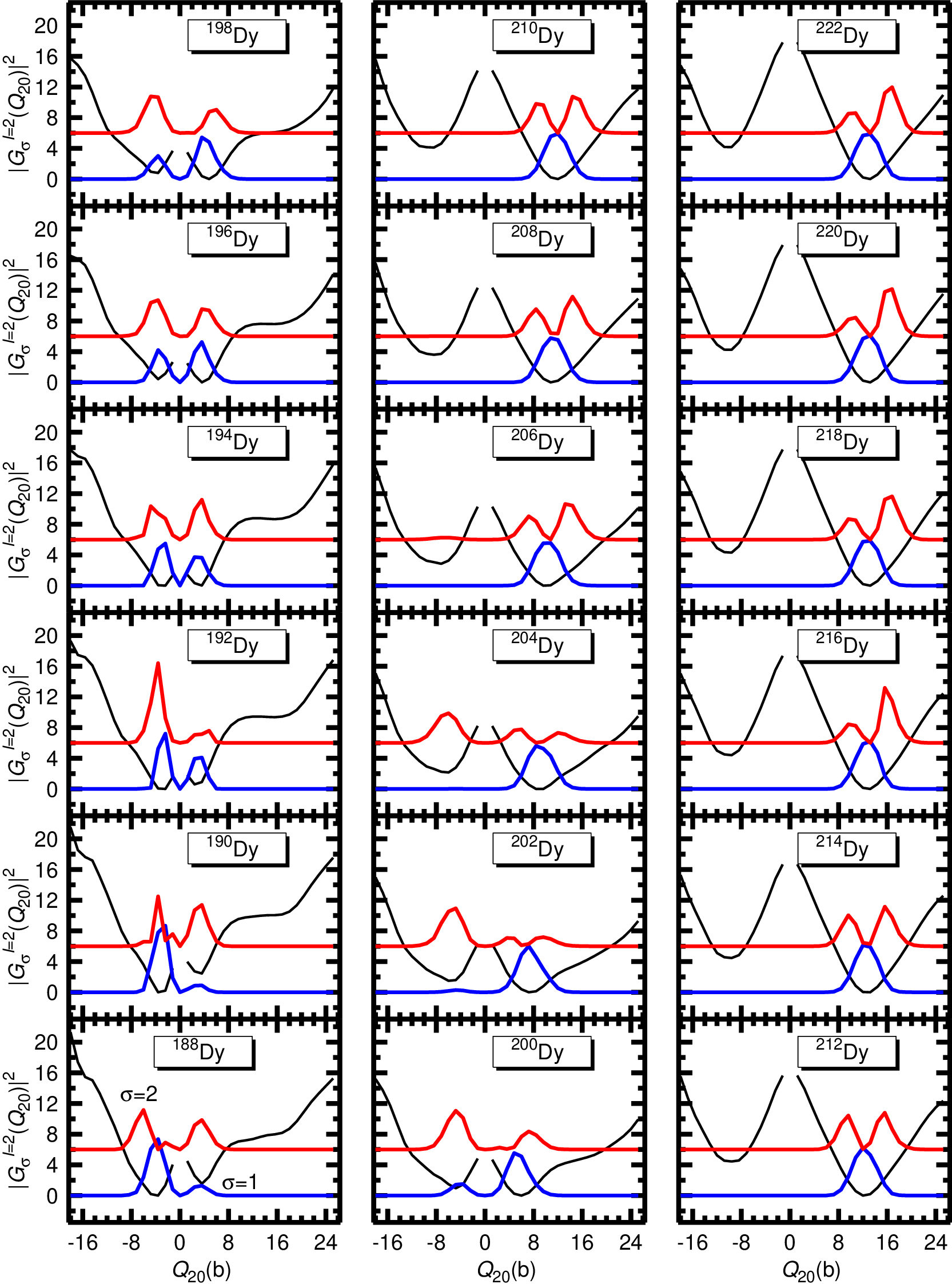}
\caption{ (Color online)
The square of the I$^{\pi}$=2$^{+}$ collective wave functions [Eq.(\ref{COLLWFS})] 
corresponding to the ground ($\sigma$=1) and first excited ($\sigma$=2) states in the 
nuclei $^{188-222}$Dy
are plotted as a function of the quadrupole moment 
$Q_{20}$.  The quantities actually depicted are 25 $\times$ $|$ G$_{\sigma=1}^{I=2}$($Q_{20}$)$|^{2}$
(blue) and 6 + 25 $\times$ $|$ G$_{\sigma=2}^{I=2}$($Q_{20}$)$|^{2}$ (red). For each nucleus the I$^{\pi}$=2$^{+}$ AMPPECs
(black)
are also included in the plots. Energies are always referred to the absolute minima of the corresponding 
AMPPECs. Results have been obtained with the parametrization D1M of the Gogny-EDF.
}
\label{fig-WFs-J2-Dy} 
\end{figure*}

The sudden decline in the S$_{2N}$ values at N=128 is a manifestation 
of the strong N=126 shell closure. In going down from Z=74 (W) to Z=60 
(Nd) no dramatic reduction occurs in the shell gap 
$\Delta_{Shell}$=S$_{2N}$(Z,126)-S$_{2N}$(Z,128)  remaining strong 
enough  as one approaches or even crosses the two-neutron dripline. 
This agrees well, with the strong shell effects predicted within the 
framework of the spherical RMF approximation \cite{N126-Farhan-2006}.

%
%
\begin{figure*}
\includegraphics[width=0.95\textwidth]{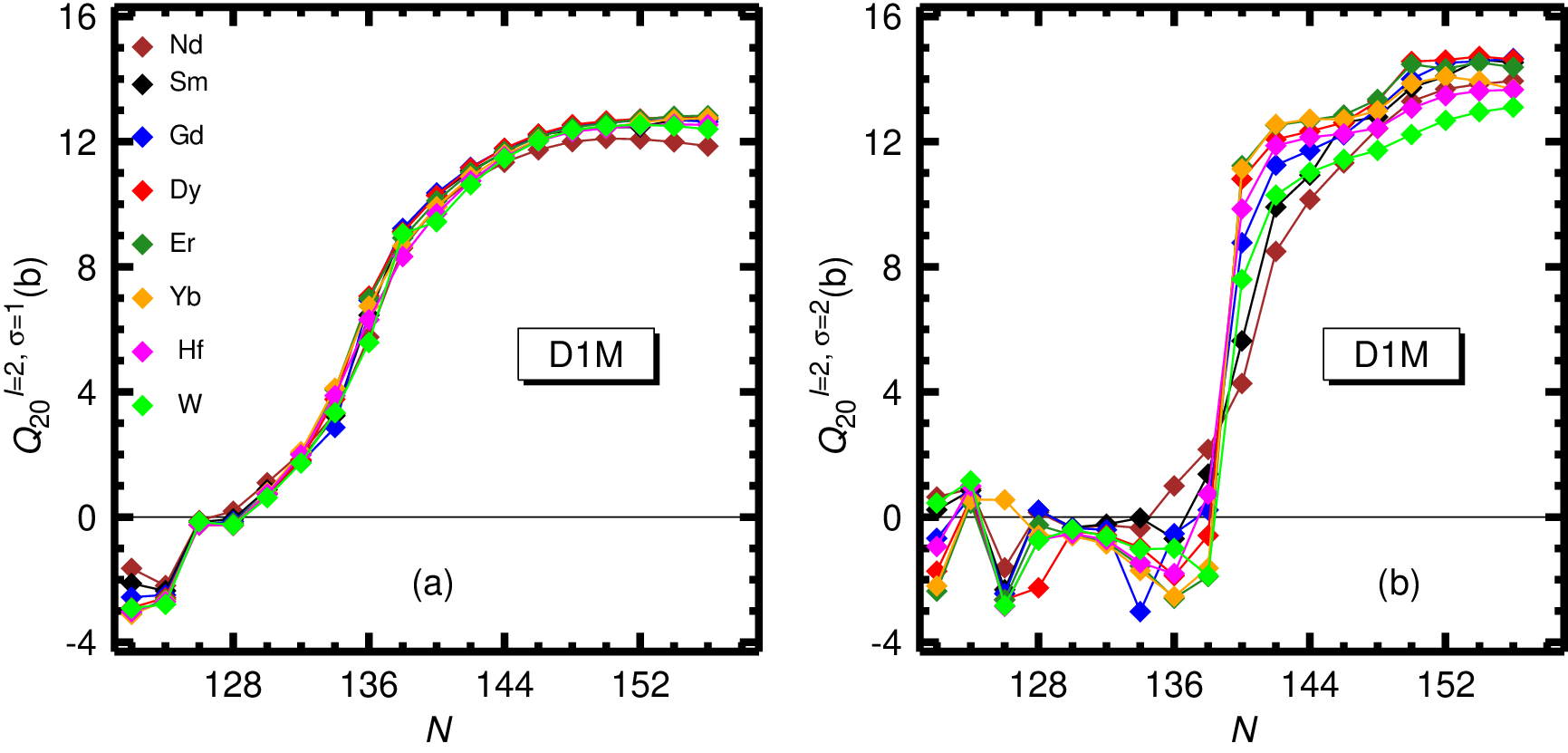}
\caption{ (Color online) The intrinsic   deformations 
[Eq.(\ref{AMPGCM-ave-Q20})]
associated with the 
ground ($\sigma$=1) and first excited  ($\sigma$=2) states  at I$^{\pi}$=2$^{+}$
in the nuclei
$^{182-216}$Nd, $^{184-218}$Sm, $^{186-220}$Gd, 
$^{188-222}$Dy, $^{190-224}$Er, $^{192-226}$Yb, $^{194-228}$Hf and $^{196-230}$W 
are depicted in panels (a) and (b), respectively, as a function of neutron number.
Results have been obtained with the parametrization D1M of the Gogny-EDF.
For more details, see the main text.
}
\label{fig-QAMPGCM-gs-summary-J2} 
\end{figure*}

As can be seen in Fig. \ref{fig-S2N-shells-D1M}, the two-neutron 
separation energies are well described in the framework of the 
spherical HFB approximation for  N=126, 128 and 130. However, the onset 
of slight oblate deformations at N=122-124 (see, Fig. 
\ref{fig-Q20-GS-summary}) leads to S$_{2N}$ values smaller than the 
ones obtained using the spherical HFB approximation. For all the 
considered isotopic chains, the difference with the spherical 
calculations becomes more dramatic beyond N=132. As we have discussed 
above, for such nuclei there is an onset of prolate ground state 
deformations (see, Fig. \ref{fig-Q20-GS-summary}) that leads to an 
additional binding energy gain. As a result, an increase in the 
two-neutron separation energies takes place in going above N=132 with a 
ridge around the neutron number N=140. In general, two-neutron 
separation energies display a decline when adding further neutrons. 
Note that, while all the Nd isotopes remain unbound beyond N=126, the 
onset of prolate ground state deformations makes the nuclei 
$^{200,202,204}$Sm stable against two-neutron decay. Furthermore, while 
in the spherical HFB approach the  two-neutron dripline is reached 
at N=132 and N=140 for the Gd and Dy isotopic chains, the presence of 
static ground state deformation effects in these chains shifts the 
location of the corresponding two-neutron driplines up to N=148 and 
N=152, respectively. For the remaining chains, the onset of ground 
state deformation enhances the  stability with respect to the spherical 
HFB results in the neutron number range between N=132 and N=156.

The question that naturally arises is to what 
extent is this enhanced stability dependent on the particular Gogny-EDF 
employed at the HFB level. In  Fig. \ref{fig-S2N-shells-D1S}, we show 
the two-neutron separation energies obtained with the Gogny-D1S EDF and 
M$_{z,Max}$=14. The observed trends resemble the ones already discussed 
for the Gogny-D1M EDF in Fig. \ref{fig-S2N-shells-D1M}. This is not 
surprising as both parametrizations provide similar HFB quadrupole 
deformation landscapes (see, Figs. \ref{fig-PEC-Sm_D1M} and \ref{fig-Q20-GS-summary}). However, as 
already anticipated, the pronounced under-binding obtained with the 
Gogny-D1S EDF leads to systematically smaller S$_{2N}$ values for 
increasing neutron number. Keeping this in mind, we conclude that the 
enhanced stability beyond the neutron shell closure N=126 is a genuine 
property of Gogny-like EDFs. This conclusion is further corroborated by resorting 
to the parametrization D1N \cite{gogny-d1n} of the Gogny-EDF which 
provides S$_{2N}$ values closer to the ones obtained with the D1M 
parameter set. Note that the results suggest, at least for some of the 
considered isotopic chains, a shift of the r-process path to higher 
neutron numbers.  

Having checked the robustness of the HFB predictions with respect to 
the considered parametrization of the Gogny-EDF, it is important to 
check their stability against quantum corrections stemming from the 
restoration of broken symmetries (mainly the rotational symmetry) and 
quadrupole configuration mixing. In addition to ground state 
properties, such beyond mean field calculations also give access to 
excited states as well as other (dynamical) quadrupole properties in 
the studied nuclei.

Regardless of the Gogny-EDF employed one observes that in the Nd, Sm, 
Gd, Dy, Er, Yb, Hf and W chains the MFPECs become wider with increasing 
neutron number and display oblate local minima. For example, we have 
found (see, Fig. \ref{fig-PEC-Sm_D1M}) that for $^{194-200}$Sm  these 
oblate local minima lie less than 2 MeV above the corresponding ground 
state. Such a shape coexistence also calls for a symmetry-projected 
configuration mixing analysis.  For larger neutron numbers, the 
excitation energy of the oblate wells increases reaching  its largest 
value (2.77 MeV and 3.87 MeV for the Gogny-D1S and Gogny-D1M EDFs, 
respectively) for $^{210}$Sm. This comes along with the development of 
spherical barriers whose height reaches 10.72 MeV (12.26 MeV) in 
$^{218}$Sm with the parametrization D1S (D1M). We stress, however, that 
even for those nuclei with well defined prolate wells it is important 
to carry out symmetry-projected configuration mixing calculations since 
the collective dynamics is determined  not only by  the energy 
landscape but also by the underlying  inertia. With this in mind 
Gogny-D1M AMPGCM calculations have been carried out (see, Sec. 
\ref{RESULTS-BYMF}) for all the nuclei studied in this paper.

Before ending this section, let us turn our attention to 
the evolution of the  SPEs in the considered nuclei
as a function of the deformation parameter $\beta_{2}$. In Fig. \ref{fig-SPE}
the proton and neutron SPEs, computed with the D1M parametrization, are
depicted for the 
N=142 nuclei $^{202}$Nd and $^{212}$Hf. Both nuclei have a prolate
deformed ground state minimum at $\beta_{2} \approx$  0.3. The first noticeable
fact is that both sets of SPEs look rather similar (up to
global energy shifts) for both nuclei in spite of the 12 units difference in proton 
number. This result shows that the gross SPE  behavior with deformation depends only on the intrinsic shape of the nucleus
and not on its proton and neutron number. It is just the position of the 
Fermi level that determines the deformation properties of the specific nucleus.
In both nuclei we observe how the level density around $\beta_{2} = 0.3$ decreases
as compared with the neighboring regions signaling the existence of the ground
state minimum at that deformation (Jahn-Teller effect). We also observe
down-slopping neutron $j_{15/2}$ orbitals plunging in the Fermi sea and
up-slopping neutron $g_{9/2}$ ones coming out. In the proton case, the down
and up-slopping orbitals are different for $^{202}$Nd and $^{214}$Hf. We
have down-slopping proton $h_{11/2}$ in the $^{202}$Nd case and $h_{9/2}$
(coming from across the Z=82 shell closure) in the  $^{214}$Hf nucleus. 
In the former nucleus the $g_{7/2}$ is the predominant up-slopping orbital whereas
in the later it is a combination of the $g_{7/2}$ (the $K=7/2$ component) and
the $d_{5/2}$.

\subsection{The AMPGCM approximation: dynamical quadrupole properties}
\label{RESULTS-BYMF}

Before turning the attention to the AMPGCM, it is illustrative to 
analyze the behavior of the AMP energies as a function of the
intrinsic quadrupole moment. In panels (a)-(d) of Fig. \ref{fig-AMPPECs-Dy}, 
the $I^{\pi}=0^{+}$  and  $2^{+}$ 
AMPPECs are shown for the nuclei $^{188,192,202,218}$Dy as a
function of the quadrupole moment $Q_{20}$. The MFPECs 
have  also been  included in the plots for comparison. All the energies 
are referred to the absolute minimum of the $I^{\pi}=0^{+}$ AMPPECs. 
The points ommited in the $I^{\pi}=2^{+}$ AMPPECs around $Q_{20}$=0 
correspond to intrinsic configurations with a very small value of the 
norm overlap  ${\cal{N}}^{I}(Q_{20},Q_{20})$ [see, Eq.(\ref{Kernels})] 
and can, therefore, be safely ommited since they do not play a role in 
the AMPGCM calculations to be discussed later on \cite{NPA-2002}. No 
energy gain is obtained due to AMP for the spherical configurations and 
$I^{\pi}=0^{+}$ since these are already pure $0^{+}$ states with 
${\cal{N}}^{I}(0,0)$=1 
\cite{NPA-2002,RaynerN20-PRC2000,RaynerN28,Rayner-Pb} .

The comparison between the $I^{\pi}=0^{+}$ AMPPECs and the MFPECs in 
both $^{188,192}$Dy  reveals the pronounced changes induced on the 
mean field energy landscapes due to the restoration of the broken 
rotational symmetry. In the case of $^{188}$Dy, the $0^{+}$ ground 
state corresponds to an oblate configuration with $Q_{20}$=-3.6 b while 
a low-lying prolate minimun ($\Delta$ E = 430 keV) is found at 
$Q_{20}$=3.6 b. These minima are separated by a spherical barrier whose 
height is 3.79 MeV. While  at the HFB level the ground state of the 
N=126 nucleus $^{192}$Dy is spherical,  once AMP is carried out two  
degenerate minima appear located at $Q_{20}=\pm 2.4$ b and separated 
by a barrier of 2.24 MeV. Such $Q_{20}$-symmetric degenerate minima 
have also been found for nuclei with spherical HFB ground states in 
several regions of the nuclear chart (see, for example, 
\cite{NPA-2002,RaynerN28,Rayner-Pb,Duguet,Bender-Pb-1}) and their 
origin can be traced back to the behavior of the rotational energy 
correction \cite{egido-Lect-Notes} near sphericity. As we will see later in 
the framework of the AMPGCM scheme, the ground state collective wave
function takes similar values around these two minima in such a way that 
the 
correlated  ground state for $^{192}$Dy is spherical. This seems to be
a general feature as the same happens for all the 
other N=126 nuclei studied in this work. For the nuclei $^{202}$Dy  and 
$^{218}$Dy, the $0^{+}$ AMPPECs exhibit well pronounced prolate minima 
at $Q_{20}$=7.2 b and 13.2 b while the oblate ones are located at 
$Q_{20}$=-4.8 b and -9.6 b. The heights of the spherical barriers are  
4.89 MeV and 16.58 MeV, respectively. In the case of the  $2^{+}$ 
AMPPECs, the absolute minima are oblate (prolate) deformed for the 
nuclei  $^{188}$Dy and $^{192}$Dy  ($^{202}$Dy and $^{218}$Dy).  Note, 
that the large excitation energy of 5.26 MeV for the oblate $2^{+}$ 
minimum ($Q_{20}$=-2.4 b) in $^{192}$Dy is consistent with the 
expectations for a spherical nucleus. The corresponding excitation 
energies for $^{202}$Dy and $^{218}$Dy  decrease to 240 keV and 105 
keV, respectively. The plots in panels (a)-(d) of Fig. 
\ref{fig-AMPPECs-Dy} already illustrate that a symmetry-projected 
configuration mixing analysis is required for the studied nuclei.

%
%
\begin{figure}
\includegraphics[width=0.45\textwidth]{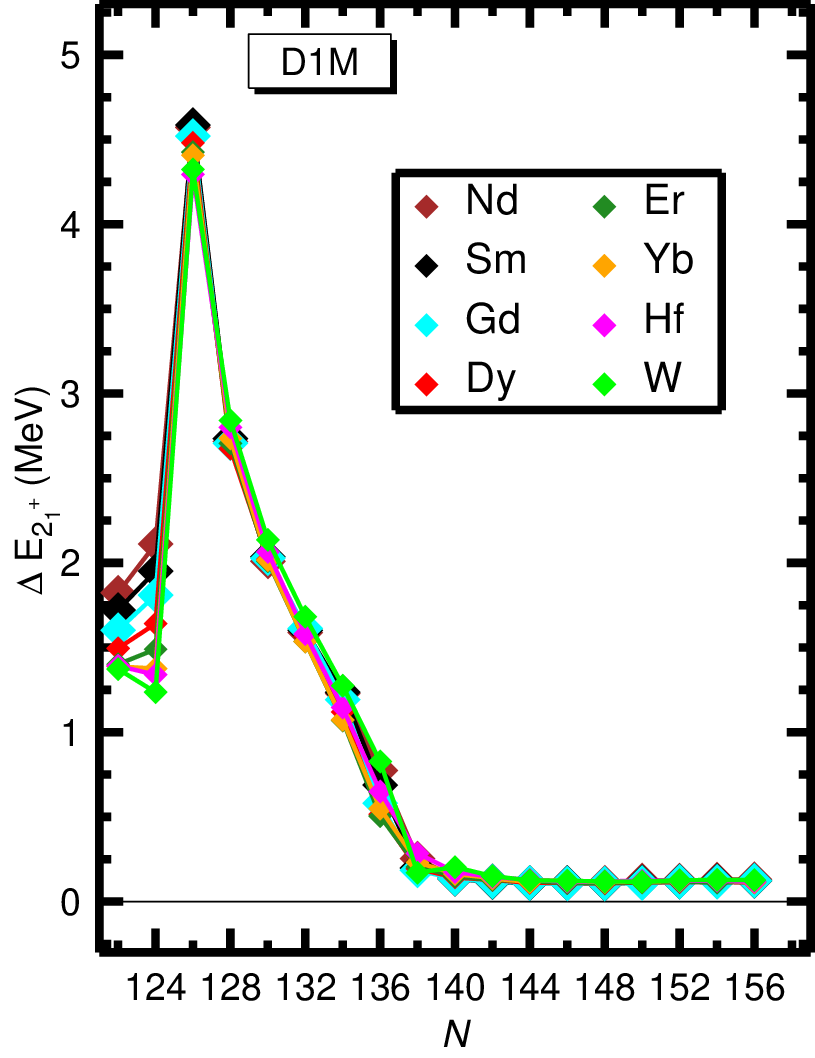}
\caption{ (Color online) The excitation energies $\Delta E_{2_{1}^{+}}= E_{2_{1}^{+}}-E_{0_{1}^{+}}$
obtained in the framework of Gogny-D1M AMPGCM calculations are plotted as a function of 
neutron number.
}
\label{fig-2plus-AMPGCM} 
\end{figure}

The square of the I$^{\pi}$=0$^{+}$ collective wave functions 
[Eq.(\ref{COLLWFS})] corresponding to the ground ($\sigma$=1) and first 
excited ($\sigma$=2) states in the nuclei $^{188-222}$Dy are plotted in 
Fig. \ref{fig-WFs-Dy} as a function of the quadrupole moment $Q_{20}$. 
To facilitate the reading of the plot, these quantities have been
stretched and shifted according to the formulas
25 $\times$ $|$ G$_{\sigma=1}^{I=0}$($Q_{20}$)$|^{2}$ and 6 + 25 $\times$ $|$ G$_{\sigma=2}^{I=0}$($Q_{20}$)$|^{2}$. 
For each nucleus the AMPPECs  are also included in the 
plots to guide the eye.

The 0$_{1}^{+}$ collective wave functions for the isotopes 
$^{188-200}$Dy exhibit a significant admixture between the prolate and 
oblate minima found in the AMPPECs. In the case of the N=126 nucleus 
$^{192}$Dy, the prolate and oblate configurations have practically the 
same weight and therefore its ground state turns out to be spherical on 
the average. The deformation effects found in the corresponding AMPPEC 
(see, Fig. \ref{fig-AMPPECs-Dy}) are not stable once quadrupole 
fluctuations are taken into account and N=126 remains, on the average 
(i.e., dynamically), as a spherical magic number not only for the Dy 
but also for all the  studied isotopic chains. On the other hand, for 
the nuclei $^{202-222}$Dy the 0$_{1}^{+}$ collective wave functions are 
well inside the prolate wells.

Having the collective wave functions $|$ 
G$_{\sigma=1}^{I=0}$($Q_{20}$)$|^{2}$ at hand, we have computed the 
average ground state deformations $Q_{20}^{I=0, \sigma=1}$ 
[Eq.(\ref{AMPGCM-ave-Q20})]. They are plotted in panel (a) of Fig. 
\ref{fig-QAMPGCM-gs-summary} as a function of neutron number. It is 
satisfying to observe that the main trend obtained within the HFB 
approximation [panel (a) of Fig. \ref{fig-Q20-GS-summary}] does survive 
the effects of zero point quantum corrections, though the AMPGCM 
calculations provide smoother shape transitions than the mean field 
ones. For all the studied isotopic chains, the AMPGCM calculations 
predict the onset of (dynamical) prolate deformations  for  N $\ge$ 
132.  

Coming back to Fig. \ref{fig-WFs-Dy}, one sees that for the heavier Dy 
isotopes, the 0$_{2}^{+}$ collective wave functions display a behavior 
reminiscent of a $\beta$-vibrational band, i.e., they are located 
inside the prolate wells and have a node at a $Q_{20}$ deformation 
where the 0$_{1}^{+}$ ground state wave functions  attain their maximum 
value. The same pattern is also observed in the Nd, Sm, Gd, Er, Yb, Hf 
and W chains. However, at least in some nuclei, the 0$_{2}^{+}$ states 
are not symmetric around the node and, therefore, they cannot be 
considered as pure $\beta$-vibrations. The average deformations 
$Q_{20}^{I=0, \sigma=2}$ [Eq.(\ref{AMPGCM-ave-Q20})] corresponding to 
the 0$_{2}^{+}$ states have also been plotted in panel (b) of 
Fig.\ref{fig-QAMPGCM-gs-summary}.  There is an onset of large prolate 
deformations in the 0$_{2}^{+}$ states around N=140-142. On the other 
hand, several shape transitions take place for N $\le$ 138. Though the 
overall trends are very similar, the precise location and nature of 
such shape transitions depend on the considered isotopic chain. 
 
The square of the I$^{\pi}$=2$^{+}$ collective wave functions for the 
ground ($\sigma$=1) and first excited ($\sigma$=2) states in 
$^{188-222}$Dy is plotted in Fig. \ref{fig-WFs-J2-Dy}. For both 
$^{188,190}$Dy the 2$_{1}^{+}$ states exhibit peaks on the oblate 
sector while for  $^{192-198}$Dy there is an  admixture of prolate and 
oblate configurations in those states. On the other hand, for the 
neutron numbers N $\ge$ 134 the 2$_{1}^{+}$ collective wave functions 
are well inside the prolate wells. The average deformations 
$Q_{20}^{I=2, \sigma=1}$ obtained for the Nd, Sm, Gd, Dy, Er, Yb, Hf 
and W chains are summarized in panel (a) of 
Fig. \ref{fig-QAMPGCM-gs-summary-J2}. In this case,  the AMPGCM 
calculations predict a transition to prolate deformed 2$_{1}^{+}$  
states for N $\ge$ 130.

Coming back to Fig. \ref{fig-WFs-J2-Dy}, the 2$_{2}^{+}$ collective 
wave functions, as well as the ones obtained for other isotopic chains, 
display a transition to a quasi-$\beta$ vibrational regime  around the 
neutron number N=140. The corresponding $Q_{20}^{I=2, \sigma=2}$ values 
have been plotted in panel (b) of Fig. \ref{fig-QAMPGCM-gs-summary-J2}. 
Besides the shape transitions observed for lighter nuclei,  there is an 
onset of large prolate deformations in the 2$_{2}^{+}$ states around 
N=140-142. The previous results indicate that, for all the studied 
nuclei, the AMPGCM zero point quantum corrections lead to predominant 
dynamical prolate deformations in the 0$_{1}^{+}$, 0$_{2}^{+}$, 
2$_{1}^{+}$ and 2$_{2}^{+}$ states with increasing neutron number.

The excitation energies of the 2$_{1}^{+}$ states are plotted in Fig. 
\ref{fig-2plus-AMPGCM} as a function of neutron number. The first 
noticeable feature is the pronounced peak at the 
neutron shell closure N=126. On the other hand, the decrease of the 
excitation energies observed for N $\ge$ 132 is well correlated with 
the onset of prolate deformations found for both the 0$_{1}^{+}$ and 
2$_{1}^{+}$ states in the AMPGCM calculations. Moreover, the values of 
such excitation energies remain small ($\Delta E_{2_{1}^{+}} <$ 300 
keV) and almost constant for nuclei with neutron numbers  N $\ge$ 
140. This is precisely the neutron sector for which large prolate 
deformations are stabilized by the AMPGCM zero point quantum 
corrections  [see, panel (a) of Figs. \ref{fig-QAMPGCM-gs-summary} and 
\ref{fig-QAMPGCM-gs-summary-J2}]. 

%
%
\begin{figure}
\includegraphics[width=0.45\textwidth]{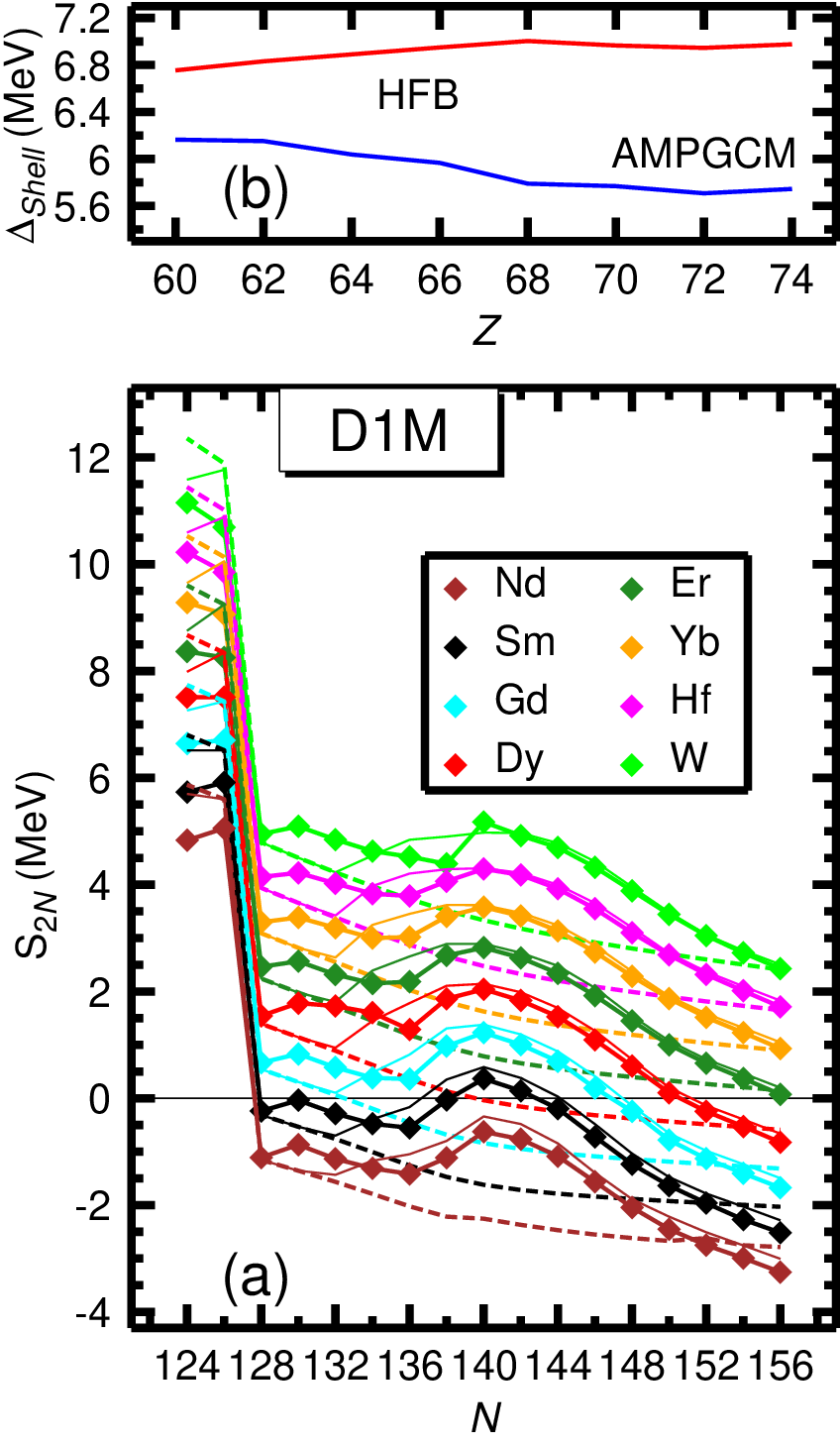}
\caption{ (Color online) The S$_{2N}$ values
computed using the energies corresponding to the 
0$_{1}^{+}$ AMPGCM ground states are depicted 
in panel (a)
(thick continuous lines with diamonds)
for the nuclei $^{182-216}$Nd, $^{184-218}$Sm, $^{186-220}$Gd, $^{188-222}$Dy, $^{190-224}$Er, $^{192-226}$Yb,
$^{194-228}$Hf and $^{196-230}$W as a function of 
neutron number. The two-neutron separation energies obtained
within the HFB framework (thin continuous lines) 
and the ones corresponding to spherical calculations (dashed lines)
are also included in the plot for comparison.
The AMPGCM (blue curve) and HFB (red curve)  shell 
gaps are plotted in panel (b) as a function
of proton number.
All the results are obtained with the D1M parametrization. 
For more details, see the main text.
}
\label{fig-S2N-AMPGCM} 
\end{figure}

In panel (a) of Fig. \ref{fig-S2N-AMPGCM} we have plotted (thick 
continuous lines with diamonds) the S$_{2N}$ values, computed using the 
energies corresponding to the 0$_{1}^{+}$ AMPGCM ground states, as a 
function of neutron number. The two-neutron separation energies 
obtained within the HFB framework (thin continuous lines) and the ones 
corresponding to spherical calculations (dashed lines) are also 
included in the plot for comparison. It is satisfying to observe that 
the trends in the AMPGCM S$_{2N}$ values support the enhanced stability 
predicted at the HFB level with respect to the spherical calculations. 
In particular, the ridge predicted at the mean field level around N=140 
does survives the effects of symmetry-projected configuration mixing. 
However, the comparison between the AMPGCM and the HFB S$_{2N}$ 
energies also reveals that while in the latter  an enhancement only 
takes place  beyond N=132 in the former it already occurs two mass 
units before.  In addition, for all the considered isotopic chains, the 
AMPGCM two-neutron separation energies at N=126 are significantly 
smaller than the HFB ones while the corresponding values at N=128 are 
almost identical. This, as can be seen from panel (b), leads to smaller 
AMPGCM shell gaps. However we stress that, in spite of this dynamical 
reduction, the shell gap remains strong enough when moving down from 
Z=74 to Z=60. The quenched AMPGCM shell gap, as compared with the HFB 
one, and its smooth decrease  with increasing Z values  agree well with the results 
obtained  for N=126 isotones in a previous (beyond mean field) global 
study of quadrupole correlation effects \cite{Bender-global-Q20}.

\section{Conclusions}
\label{conclusions}

In this work, we have considered the behavior of quadrupole 
collectivity across the N=126 neutron shell closure in the Nd, Sm, Gd, 
Dy, Er, Yb, Hf and W isotopic chains including very neutron-rich nuclei 
up to N=156. In the HFB framework N=126 nuclei are found to have 
spherical ground states and the corresponding mean field shell gaps do 
not exhibit a dramatic reduction when approaching or even crossing  the 
two-neutron driplines.

Spherical HFB calculations are inappropriate to describe the static 
quadrupole properties of nuclei in the  vecinity of the neutron shell 
closure. In particular, a shape transition to prolate deformed ground 
states is predicted to occur around N=132-134 within the (constrained) 
HFB framework. As a consequence of the onset of static axially 
symmetric quadrupole deformations there is an enhancement of the 
two-neutron separation energies that extends the corresponding 
two-neutron driplines far beyond what could be expected within 
spherical calculations. We have shown that such an enhanced stability 
is a genuine property of Gogny-like EDFs, i.e., it is independent of 
the particular parametrization employed in the calculations. The 
analysis  of the SPEs reveals the presence of  Jahn-Teller distorsions 
in the corresponding proton and neutron spectra associated with the 
global and local minima found in the MFPECs.

Moreover, the constrained HFB calculations reveal that some  nuclei 
display shape coexistence, i.e., low-lying prolate and oblate minima 
with similar energies. Such a shape coexistence also shows up in the 
AMPPECs and calls for a symmetry-projected configuration mixing 
analysis. Within this context, beyond mean field correlations, stemming 
from the restoration of the rotational symmetry broken at the HFB level 
and quadrupole fluctuations, have been taken into account, for all the 
considered Nd, Sm, Gd, Dy, Er, Yb, Hf and W nuclei, in the framework of 
the AMPGCM scheme based on the Gogny-D1M EDF.

Our AMPGCM calculations provide a dynamically correlated spherical 
ground state for all the studied N=126 nuclei. The  AMPGCM correlations 
induce a reduction of the N=126 two-neutron shell gap as compared with 
the mean field one, in good agreement with other calculations 
\cite{Bender-global-Q20}. However, we stress that on their own the 
AMPGCM shell gaps remain strong enough all the way down from Z=74 (W) 
to Z=60 (Nd). 
 
Both the wave functions and average deformations obtained within the 
AMPGCM framework indicate that, with increasing neutron number, beyond 
mean field zero point quantum corrections stabilize dominant prolate 
configurations not only in the 0$_{1}^{+}$ but also in the 0$_{2}^{+}$, 
2$_{1}^{+}$ and 2$_{2}^{+}$ collective states. The dynamical onset of 
large deformations along the considered isotopic chains is well 
correlated, for example, with the behavior of the 2$_{1}^{+}$ 
excitation energies as a function of neutron number. On the other hand, 
for the heavier nuclei the 0$_{2}^{+}$ and 2$_{2}^{+}$ collective wave 
functions exhibit a behavior reminiscent of $\beta$-vibrational bands. 
 
It is found that, as a consequence of the onset of dynamical quadrupole 
deformations in the 0$_{1}^{+}$ states, the computed AMPGCM two-neutron 
separation energies corroborate the enhanced stability predicted at the 
mean field level. In particular, we have shown that the sudden spur in 
the two neutron separation energies, with a ridge around N=140, does 
survive the effects of zero point quantum fluctuations. Within this 
context, at least for some of the studied isotopic chains, the AMPGCM 
correlations shift the occurrence  of the two-neutron driplines to 
higher neutron numbers.

We believe that the results discussed in this work deserve further 
attention not only from the nuclear structure side but also due to 
their possible consequences for the r-process path around N=126. A long 
list of task remains to be undertaken. For example, we have shown the 
indepedence of our HFB predictions with respect to the Gogny-EDF 
employed in the calculations. It would be interesting to compare with 
the mean field and beyond mean field predictions arising from other 
state-of-the-art relativistic and nonrelativistic approaches. On the 
other hand, a more realistic treatment of pairing correlations, 
including a symmetry-projected analysis of the coupling between pairing 
and quadrupole degrees of freedom, is left for a future study. Work 
along these lines is in progress and will be published elsewhere.

\begin{acknowledgments}
The work of L. M. Robledo has been supported in part
by the Spanish MINECO Grants No. FPA2012-34694, and No. FIS2012-34479
and by the Consolider-Ingenio 2010 Program MULTIDARK CSD2009-00064.
\end{acknowledgments}

\end{document}